\documentclass[aps,english,prx,floatfix,amsmath,superscriptaddress,tightenlines,twocolumn]{revtex4-2}
\usepackage[pdftex,colorlinks=true,linkcolor=darkblue,citecolor=blue,urlcolor=darkred]{hyperref}
\usepackage[english]{babel}
\usepackage{braket}
\usepackage{amsmath,amsfonts, amssymb, amsthm, dsfont}
\usepackage{yfonts,bbm}
\usepackage{bm,breqn}
\usepackage{mathrsfs}
\usepackage{graphicx}
\usepackage[caption=false]{subfig}
\usepackage[export]{adjustbox}
\usepackage[normalem]{ulem}
\usepackage{verbatim}
\usepackage{xcolor}

\usepackage{float}

\theoremstyle{definition}

\definecolor{darkblue}{rgb}{0.,0.,0.4}
\definecolor{darkred}{rgb}{0.5,0.,0.}

\usepackage{verbatim}
\usepackage{stackengine}

\usepackage{tikz}
\usetikzlibrary{calc}
\usepackage{multirow}
\usepackage[capitalise]{cleveref}

\newcommand{\calE}{\mathcal{E}}

\newcommand{\calG}{\mathcal{G}}
\newcommand{\calL}{\mathcal{L}}
\newcommand{\calC}{\mathcal{C}}
\newcommand{\calP}{\mathcal{P}}
\newcommand{\calB}{\mathcal{B}}
\newcommand{\calH}{\mathcal{H}}
\newcommand{\calT}{\mathcal{T}}

\DeclareMathSymbol{:}{\mathord}{operators}{"3A}

\renewcommand{\tilde}{\widetilde}
\renewcommand{\d}{\mathrm{d}}
\newcommand{\bs}{\textbf{s}}
\renewcommand{\bm}{\textbf{m}}
\newcommand{\be}{\textbf{e}}

\newcommand{\bc}{\textbf{c}}
\renewcommand{\mod}{~\textrm{mod}~}

\newcommand{\tr}{\text{tr}}
\renewcommand{\Pr}{\mathrm{Pr}}

\newcommand{\polylog}{{\rm polylog}}

\newcommand{\eqfig}[2]{\vcenter{\hbox{\includegraphics[height=#1]{#2}}}}

\begin{document}
\title{Stability of mixed-state quantum phases via finite Markov length}

\author{Shengqi Sang}
\affiliation{\PI}
\affiliation{\UW}
\affiliation{\KITP}

\author{Timothy H. Hsieh}
\affiliation{\PI}

\newcommand*{\PI}{Perimeter Institute for Theoretical Physics, Waterloo, Ontario N2L 2Y5, Canada}
\newcommand*{\UW}{Department of Physics and Astronomy, University of Waterloo, Waterloo, Ontario N2L 3G1, Canada}
\newcommand*{\KITP}{Kavli Institute for Theoretical Physics, University of California, Santa Barbara, CA 93106, USA}

\begin{abstract}
    For quantum phases of Hamiltonian ground states, the energy gap plays a central role in ensuring the stability of the phase as long as the gap remains finite. We propose Markov length, the length scale at which the quantum conditional mutual information (CMI) decays exponentially, as an equally essential quantity characterizing mixed-state phases and transitions. For a state evolving under a local Lindbladian, we argue that if its Markov length remains finite along the evolution, then it remains in the same phase, meaning there exists another quasi-local Lindbladian evolution that can reverse the former one. We apply this diagnostic to toric code subject to decoherence and show that the Markov length is finite everywhere except at its decodability transition, at which it diverges. CMI in this case can be mapped to the free energy cost of point defects in the random bond Ising model. This implies that the mixed state phase transition coincides with the decodability transition and also suggests a quasi-local decoding channel.
\end{abstract}
\maketitle
\emph{Introduction---}
The advent of controllable open quantum systems, whereby tailored noise and dissipation or measurement and feedback can evade thermalization, has enabled a plethora of non-equilibrium phenomena in mixed states~\cite{diss1, diss2, hastings2011nonzero, coser2019classification, fan2023diagnostics, bao2023mixed, lee2023quantum, zou2023channeling, de2022symmetry, ma2023average, zhang2022strange, ma2023topological, rakovszky2024defining, sang2023mixed, lu2023mixed, lee2022decoding, zhu2022nishimoris, chen2023symmetryenforced, chen2023separability, lessa2024mixedstate, chen2024unconventional, lee2022symmetry, lee2024exact, sohal2024noisy, ellison2024towards, wang2023intrinsic, wang2023topologically, wang2024anomaly, xue2024tensor, guo2024locally, ma2024symmetry}.  For example, certain types of noise can preserve and in some cases even enrich long-range entanglement in pure states~\cite{fan2023diagnostics, bao2023mixed, lee2023quantum, zou2023channeling}, and measurement and feedback can convert short-range entangled pure states into long-range entangled mixed states~\cite{lu2023mixed, lee2022decoding, zhu2022nishimoris, li2021robust}. These phenomena raise a fundamental question: under what circumstances do such states constitute phases of matter?

For pure states, two states are in the same phase if they are related by local unitary evolution. Such an operational characterization provides a powerful tool for identifying universal properties, \textit{i.e.} properties shared by all states within a phase. 
For mixed states, an analogous operational definition~\cite{coser2019classification} is that two mixed-states are in the same phase if they can be converted into each other through local Lindbladian evolution. Unlike the pure state case, here finding one direction of evolution does not imply the other direction because Lindbladian evolution is generally non-invertible; thus the two connections need to be obtained separately. For this purpose, a real-space renormalization scheme for mixed states and local versions of decoders were developed to provide a unified approach for navigating mixed state quantum phases~\cite{sang2023mixed}. 

Yet there remain several fundamental questions concerning mixed-state quantum phases. For pure states, a natural class of physical states is ground states of local Hamiltonians with an energy gap.  The gap is not only a defining property of pure state quantum phases but also controls other key properties: exponential decay of correlations~\cite{hastings2006spectral} and stability under Hamiltonian perturbation~\cite{bravyi2010topological, michalakis2013stability}. If the gap remains finite under local Hamiltonian perturbations, then the perturbed ground states are guaranteed to be in the same phase.  For mixed states, which are not necessarily associated with any Hamiltonian, what is the analogue of the gap? 
We would like a criteria for the stability of a mixed-state phase under physical perturbations: if it is fulfilled by the evolving state along a short-time Lindbladian evolution, all states along the path should be in the same phase.  Ideally such a criteria specifies a class of `physical' mixed states, which should include Gibbs states and local Lindbladian steady states. 
(We remark that there are different kinds of perturbations when considering stability; for example, \cite{cubitt2015stability, rakovszky2024defining, liu2024dissipative} consider the stability of a Lindbladian steady state's phase with respect to perturbing the Lindbladian.)



\begin{figure}[t]
    \centering
    \includegraphics[width=0.95\linewidth]{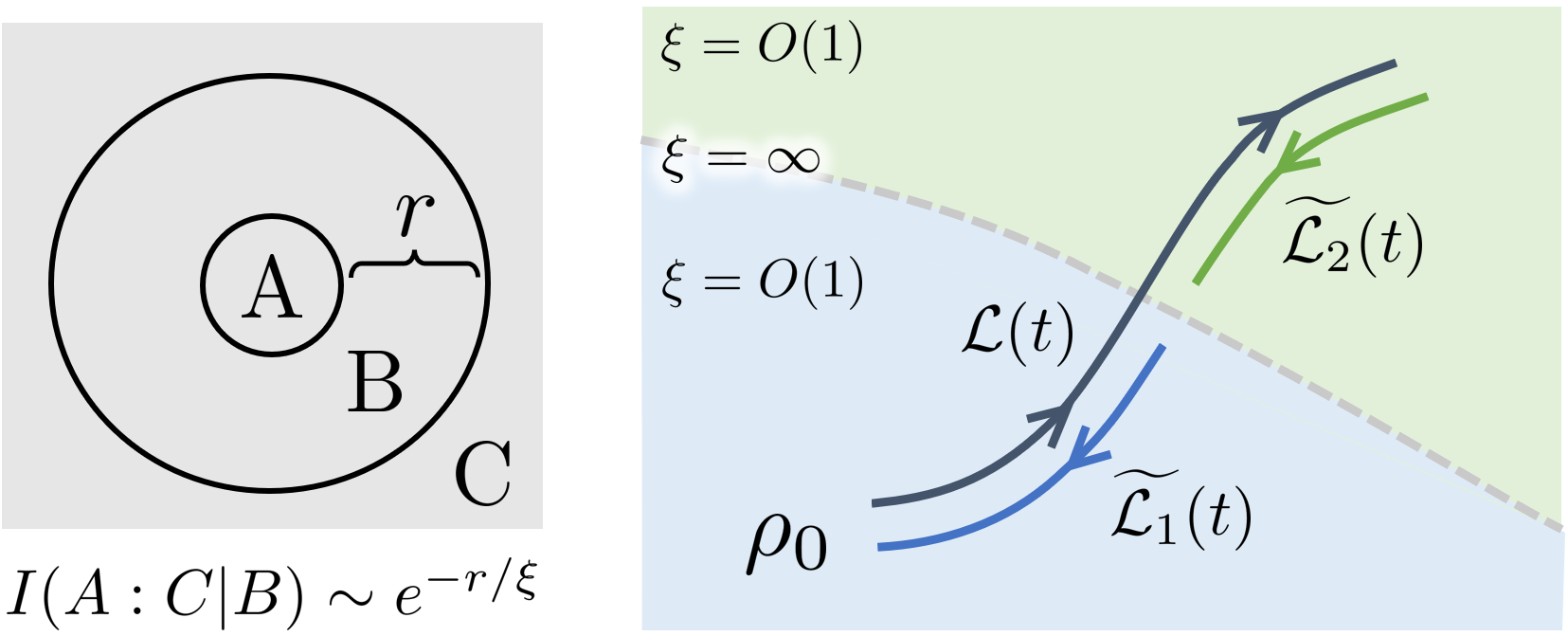}
    \caption{
    \textbf{(left)} Quantum conditional mutual information $I(A:C|B)$ quantifies how non-local is the correlation between $A$ and its complement. When $I(A:C|B)$ decays exponentially with $B$'s width $r$, we call the corresponding length scale the Markov length $\xi$.
    \textbf{(right)} The dark grey line is a path of mixed-states generated from local Lindbladian evolution, \textit{i.e.} $\rho_t=\calT\exp(\int_0^t\calL(\tau)\d\tau)[\rho_0]$. For each segment in which $\xi$ remains finite, $e.g.$ below (above) the dashed line, we argue there exists a quasi-local Lindbladian $\tilde{\calL}_{1(2)}$ that reverses $\calL$'s action. Thus states within each segment are in the same mixed-state phase.
    }
    \label{fig: main}
\end{figure}
In this work, we show that the scaling of quantum conditional mutual information (CMI)
defined as $I(A:C|B) = S(AB) + S(BC) - S(B) - S(ABC)$ for a tripartition (Fig.1) is closely tied to the reversibility of Lindbladian evolution and provides the desired criteria in the above questions. CMI (for different partitions) has provided insight into ground state topological phases~\cite{PhysRevLett.96.110404,PhysRevLett.96.110405,SHI2020168164}, quantum dynamics involving measurement and noise~\cite{sanghybrid, zhang2024nonlocal, lee2024universal}, and the efficiency of preparing and learning quantum states~\cite{onorati2023efficient, brandao2019finite, gondolf2024conditional}. Here we use it to understand mixed state phases of matter.
To gain some intuition, CMI can be written as a difference of mutual informations $I(A{:}C|B)=I(A{:}BC)-I(A{:}B)$, and thus small CMI implies  that $A$'s correlation with its complement is mostly captured by a buffer region $B$ surrounding it. For a large variety of non-critical pure and mixed states, CMI decays exponentially with the separation between $A$ and $C$, \textit{i.e.} ${\rm CMI}\sim\exp(-{\rm dist}(A,C)/\xi)$, with a length scale $\xi$ that we call the Markov length.   

Our main finding is that if the Markov length $\xi$ remains finite when evolving a state $\rho$ with a local Lindbladian, then all states along the evolution are guaranteed to be in the same mixed-state phase. More specifically, utilizing the approximate Petz theorem~\cite{junge2018universal}, we argue that there exists another  (quasi-)local Lindbladian evolution that can reverse the original evolution, and the locality of the reverse evolution is set by the maximal $\xi$ along the path. 
Thus, for mixed states the finite Markov length plays the role of and encompasses the finite energy gap (and corresponding finite correlation length) for ground states phases. (For a pure state, $I(A:C|B)=I(A:C)$, which for a gapped ground state is expected to decay exponentially with ${\rm dist}(A,C)$.) 

We apply our general result to a concrete example, the dephased toric code state. 
Viewed as a quantum error correcting code, the toric code undergoes a decodability transition~\cite{dennis2002topological} at a finite dephasing strength $p_c$ that can be mapped to the random bond Ising model (RBIM)'s ferromagnet to paramagnet phase transition.  Recently, the transition has been characterized from the perspective of mixed state properties, including topological entanglement negativity~\cite{fan2023diagnostics, neg} and many-body separability~\cite{chen2023separability}. 
Nevertheless, it has not been clear if the states before and after $p_c$ constitute two mixed state phases under the definition of~\cite{coser2019classification}. 
Also, unlike conventional critical points, the critical mixed state at $p_c$ does not have power-law decaying correlation functions, making it unclear what is the physical diverging length scale at this critical point.
We answer both questions via the Markov length defined from CMI. We find that both above and below $p_c$, the Markov length $\xi$ is finite. From our general result, we conclude the small $p$ phase is in the same phase as pure toric code, and the high $p$ phase is topologically trivial. At the critical point $p_c$, the Markov length diverges and CMI has power-law decay.  By identifying the Markov length with the RBIM correlation length, we strengthen the connection between mixed state phase and decodability transitions.

\emph{Setup and definitions---}
We start by giving a formal definition of mixed-state phase equivalence: two states $\rho_1$ and $\rho_2$ defined on a lattice of linear size $L$ are in the same phase if there exist two time-dependent quasi-local Lindbladian evolutions $\calL^1(\tau)$, $\calL^2(\tau)$ such that:
\begin{equation}
    \left|\calT e^{\int_0^1 \calL^{1,2}(\tau)\d \tau}[\rho_{1,2}]-\rho_{2,1}\right|_1 \leq \epsilon
\end{equation}
Here $\calL(\tau)=\sum_x \calL_x(\tau)$ being quasi-local means each $\calL_x(\tau)$ has $O(\polylog L)$ spatial support and $O(\polylog L)$ operator norm at any time. This definition is a technically simplified version of the one introduced in~\cite{coser2019classification}.

Next we introduce Markov length, the key quantity in this work. Let $\rho$ be a state defined on a $D$-dimensional lattice. We say $\rho$ has Markov length $\xi$ if its conditional mutual information (CMI) satisfies
\begin{equation}
    I_{\rho}(A:C|B) \leq {\rm poly}(|A|, |C|) e^{-{\rm dist}(A, C)/\xi}
\end{equation}
for any three regions $A, B, C$ with topology displayed in Fig.\ref{fig: main}(a), namely $A$ is simply connected, $B$ is an annulus-shaped region surrounding $A$, and $C=\overline{A\cup B}$ is the rest of the system. When $\rho$ can be consistently defined on lattices of arbitrarily large size $L$, 
and if $\xi$ is independent of $L$, we say the state has $\xi$-finite-Markov-length ($\xi$-FML). 


\emph{Reversing a local quantum operation---}
Assume a quantum channel $\calE$ acts on a local region $A$ of state $\rho$, and let $B$ be a width-$r$ annulus region surrounding $A$. Can the effect of $\calE$ be approximately reversed by another quantum channel $\tilde{\calE}$ acting on the enlarged region $A\cup B$? 

Using the approximate data processing inequality~\cite{junge2018universal}, we prove there exists a map $\tilde{\calE}$ acting on $A\cup B$ whose recovery quality satisfies (see SM for a derivation) \footnote{All the logarithms in this work, including those show up in the definition of entropic quantities, use $2$ as the base.}:
\begin{align}\label{eq: single_err_bound}
    {\textstyle \frac{1}{2\ln2}}
    |\tilde{\calE}\circ\calE[\rho] - \rho|_1^2 \nonumber
    &\leq I_\rho(A:C|B) - I_{\calE[\rho]}(A:C|B)\\
    &\leq I_\rho(A:C|B)
\end{align}
where the second inequality follows from strong subadditivity and $|\sigma|_1\equiv\tr(\sqrt{\sigma^\dagger\sigma})$ is the trace norm. The explicit form of $\tilde{\calE}$ is given by the twirled Petz recovery map $\tilde{\calE}=\calP(\calE, \rho_{AB})$ (see SM for the explicit expression), which depends on the forward channel $\mathcal E$ and the local reduced density matrix $\rho_{AB}$. 
The above bound shows that if $\rho$ has $\xi$-FML, local recovery is possible because the recovery error $|\tilde{\calE}\circ\calE[\rho]-\rho|_1$ decays exponentially with $r$, the width of $B$.

\emph{Reversing a continuous evolution--- } Now we turn to the local reversibility of a continuous time evolution acting everywhere on an FML state. Generically, such dynamics can be represented as a time-dependent local Lindbladian evolution: $\calG\equiv \calT e^{\int_{0}^{1}\calL_\tau\d\tau}$ where $\calL_t\equiv\sum_{x}\calL_{t,x}$ is a local Lindbaldian at all time. But for technical simplicity, we consider a discretized (or `Trotterized') Lindbladian dynamics:
\begin{align}\label{eq: forward_circuit}
    \calG &\equiv\calC_{\ell=\delta t^{-1}}\circ...\circ\calC_{\ell=2}\circ\calC_{\ell=1}
    \nonumber\\
    \calC_\ell &\equiv {\prod}_{x} \calE_{\ell, x}={\prod}_{x} e^{\delta t \calL_{\ell, x}}
\end{align}
Here each $\calL_{\ell, x}$ is a Lindbladian superoperator acting on a region referred to as $A_{\ell, x}$, and $x$ indexes gates within a layer and $\ell$ indexes time steps. For a given $\ell$, different $A_{\ell, x}$s are non-overlapping. Thus the total map $\calG$ takes the form of a circuit, whose gates are Lindbladian evolutions with a small time $\delta t$.

\begin{figure}[h!]
    \centering
    \includegraphics[width=0.97\linewidth]{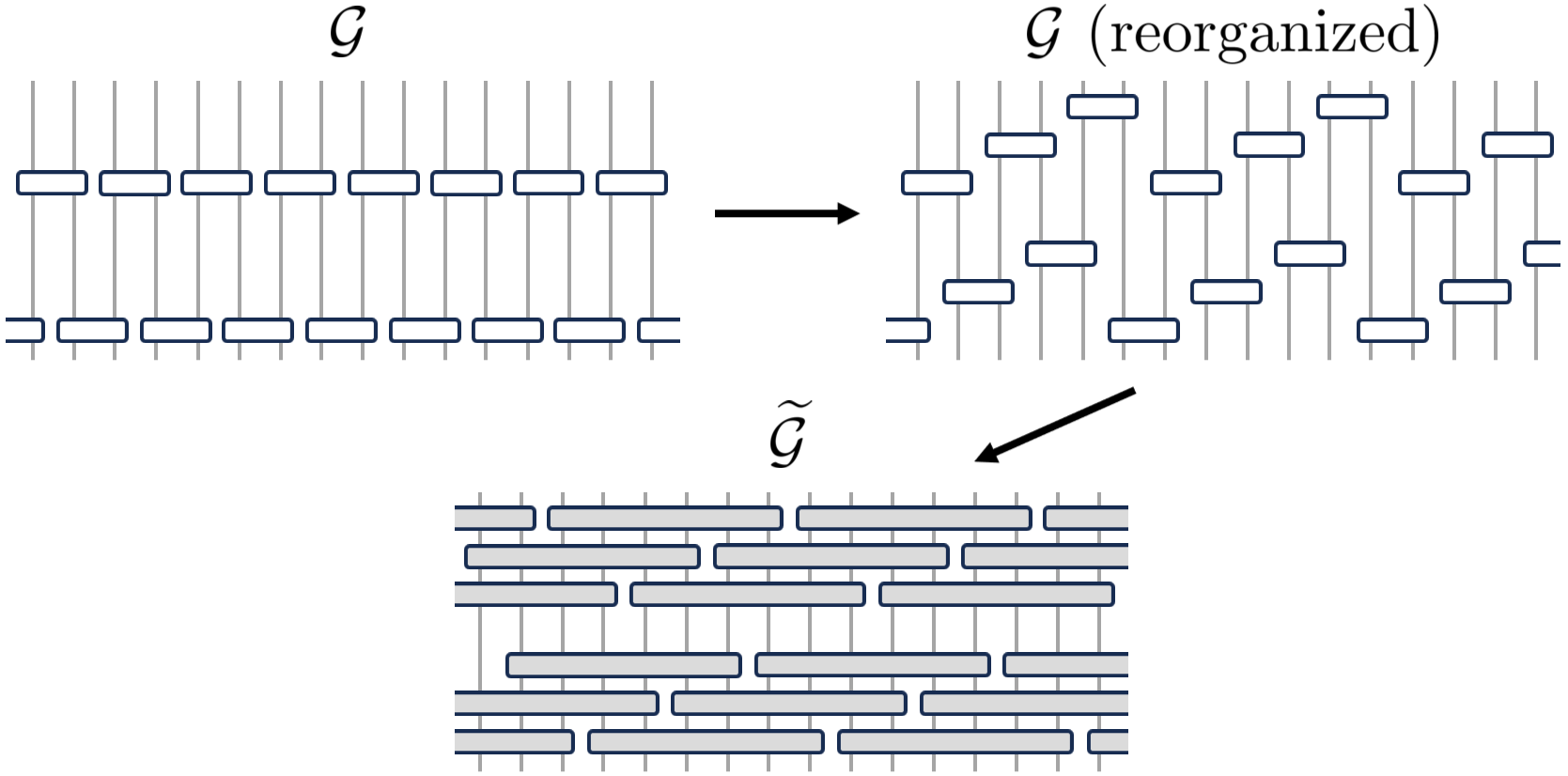}
    \caption{
   \textbf{Reversal circuit for continuous evolution---}
   (top left) Two layers of the forward circuit $\calG$ acting from bottom to top. Each box is a quantum channel $\calE_{x,t}=\exp(\delta t\calL_{x,t})$. (top right) Gates in each layer are reorganized into multiple layers so that any two gates in a layer are at least distance $2r$ separated (in the figure $r=2$).
    (bottom) The reversal circuit $\widetilde{\calG}$ constructed from the reorganized $\calG$ by replacing each $\calE_{x,t}$ with its reversal $\tilde{\calE}_{x,t}$ (grey box) defined in Eq.\eqref{eq: reversal_circuit}. The reversal circuit acts from top to bottom.}
    \label{fig: circuit}
    \vspace{-0.6cm}
\end{figure}

We show that $\calG$'s effect on $\rho$ can be reversed by another (quasi-)local evolution $\tilde{\calG}$ \textit{if $\rho$'s Markov length remains finite throughout the dynamics $\calG$}. More precisely, we assume that for any $\ell\in\{1,...,\ell_{\rm max}=\delta t^{-1}\}$:
\begin{enumerate}
    \item $\rho_\ell\equiv \calC_\ell [\rho_{\ell-1}]$ is $\xi$-FML (let $\rho_0\equiv\rho$), and further
    \item $\calC'_\ell[\rho_{\ell-1}]$ is $\xi$-FML, for $\calC'_\ell$ formed from any subset of gates within $\calC_\ell$
\end{enumerate}
The (technical) second condition is expected to follow from the first condition and small enough $\delta t$; it corresponds to the intuition that $\xi$ does not suddenly change from being finite to infinite under a small perturbation. In the absence of a proof, we leave it as an assumption for the sake of rigor.

Now we describe the reversal circuit $\tilde{\calG}$. The big picture is to reverse the circuit gate by gate using the recovery map described in the previous section in a carefully chosen order.  To this end, we first reorganize the circuit structure of $\calG$: for each layer $\calC_{\ell}$, we reorganize gates within it into multiple layers such that gates within each newly formed layer are at least distance $2r$ separated from each other ($r$ is a distance parameter we will specify later). After the reorganization, the circuit depth may increase by a factor of $O(r^D)$, but the number of gates is unchanged. Importantly, the new circuit still satisfies aforementioned conditions 1, 2 with respect to the state $\rho$ for the same $\xi$ (thanks to condition 2). With a slight abuse of notation, we still use $\{\calC_\ell\}$ to denote layers in the reorganized circuit.

The reversal dynamics can be explicitly written as:
\begin{align}\label{eq: reversal_circuit}
    \tilde{\calG} &\equiv\tilde{\calC}_{\ell=1}\circ\tilde{\calC}_{\ell=2}\circ...\circ\tilde{\calC}_{\ell=\delta t^{-1}}\\
    \tilde{\calC}_\ell &\equiv {\prod}_{x} \tilde{\calE}_{\ell, x}
    \quad
    \tilde{\calE}_{\ell, x} \equiv \calP(\calE_{\ell, x}, (\rho_{\ell-1})_{A_{\ell,x}\cup B_{\ell,x}})
    \nonumber
\end{align}
where each ${\calE}_{\ell, x}$ acts on both $A_{\ell, x}$ and a width-$r$ annulus $B_{\ell,x}$ surrounding $A_{\ell, x}$. We emphasize that each $\calE_{\ell, x}$'s reversed channel $\tilde{\calE}_{\ell, x}$ is calculated using $\rho_{\ell-1}$ as the reference state. Each $\tilde{\calE}_{\ell,x}$ also admits a (time-dependent) Lindbladian representation, i.e. $\tilde{\calE}_{\ell, x}=\calT\exp(\int_0^{\delta t} d\tau \tilde{\calL_{\ell,x}}(\tau))$, as shown in~\cite{kwon2022reversing}.
Thanks to the reorganization step, different $\tilde{\calE}$ s are non-overlapping and thus commute with each other. This guarantees, as we show rigorously in the SM, that the cumulative recovery error is bounded by the sum of single-step errors, namely:
\begin{equation}\label{eq: total_err_bound}
    \epsilon\equiv|\tilde{\calG}\circ\calG[\rho]-\rho|_1 \leq \sum_{\ell,x}
    \left|\tilde{\calE}_{\ell,x}\circ\calE_{\ell,x}\left[\rho_{\ell-1}\right] - \rho_{\ell-1}
    \right|_1\ .
\end{equation}
Since each $\rho_{\ell}$ is $\xi$-FML, according to Eq.\eqref{eq: single_err_bound} each of the terms in the \textit{r.h.s.} is bounded by ${\rm poly}(L) e^{-r/2\xi}$ with $L$ being the system size. In order to achieve the global recovery error $\epsilon$, it suffices to require
\begin{equation}
    r\geq \xi \cdot \log\left(\frac{{\rm poly}(L)}{\epsilon\cdot\delta t}\right)
\end{equation}
Thus a quasi-local \footnote{Naively the evolution time of the reversed dynamics {$\tilde{\calG}$} is {$O(r^{D})$} because the circuit depth is multiplied by the same factor due to the reorganization. However it can be turned into a time $1$ evolution by absorbing the factor into {$\calL(\tau)$}.} reversal circuit is sufficient to achieve a high recovery fidelity. Using the definition of phase equivalence, we conclude that $\calG[\rho]$ and $\rho$ are in the same phase. We observe that the conclusion does not change even if we let $\epsilon$ and $\delta t$ scale with the system size as $1/{\rm poly}(L)$. 

\begin{figure*}
    \centering
    \subfloat[]{\includegraphics[width=0.15\linewidth, height=0.22\linewidth]{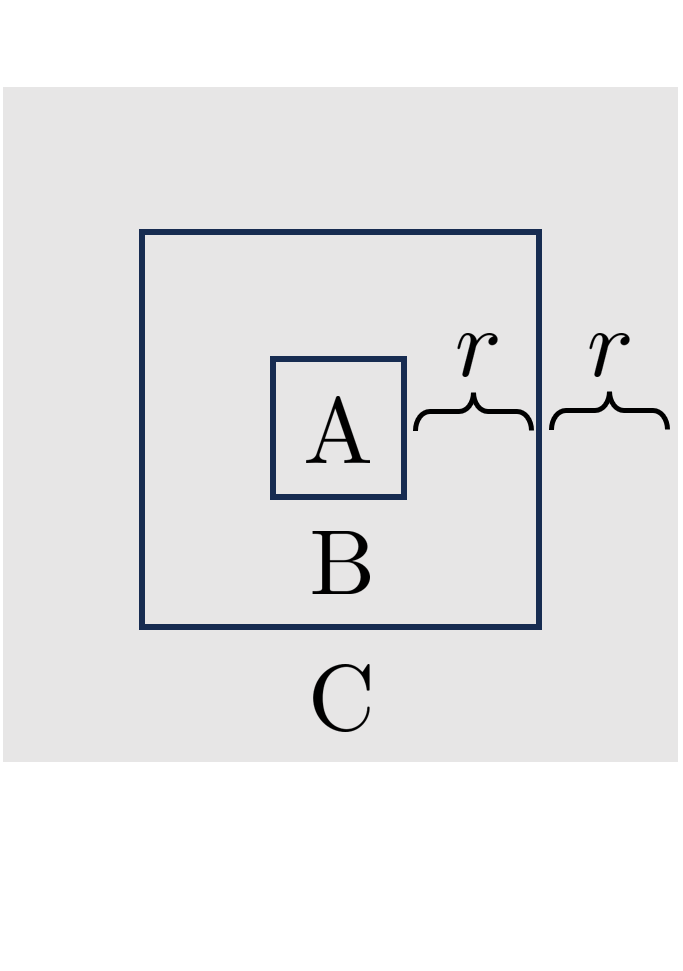}}
    \hspace{0.013\linewidth}
    \subfloat[]{\includegraphics[width=0.27\linewidth, height=0.22\linewidth]{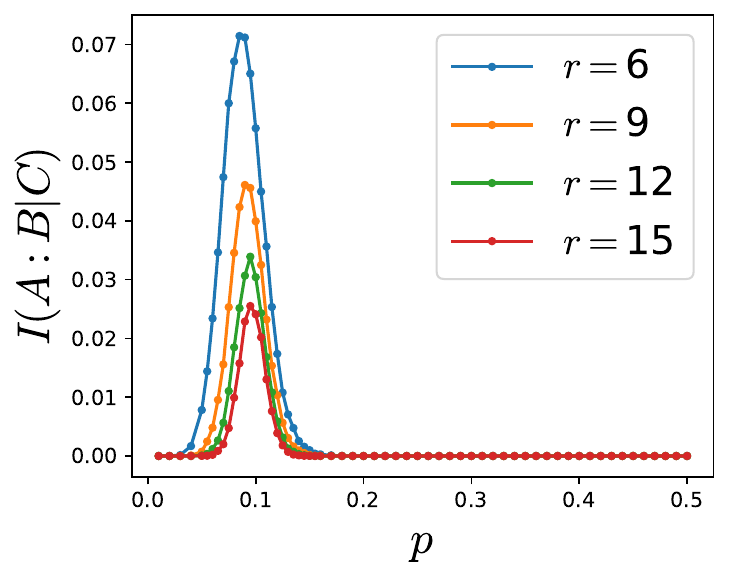}}
    \hspace{0.005\linewidth}
    \subfloat[]{\includegraphics[width=0.27\linewidth, height=0.22\linewidth]{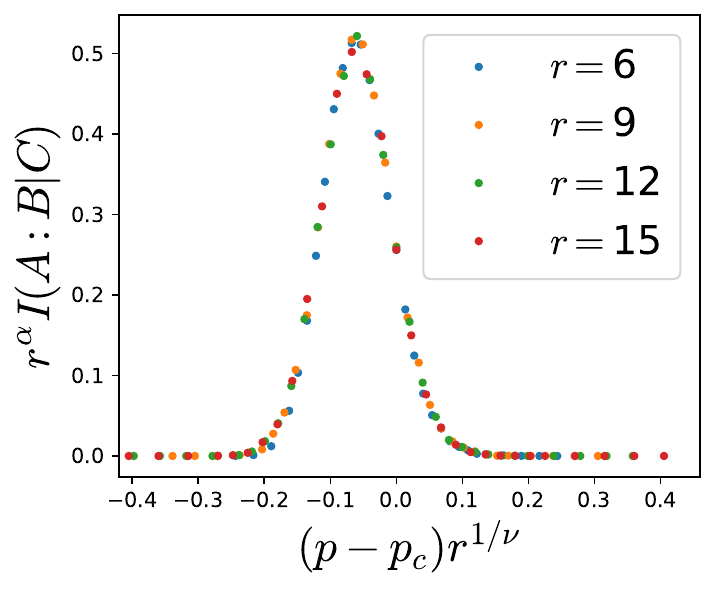}}
    \hspace{0.005\linewidth}
    \subfloat[]{\includegraphics[width=0.27\linewidth, height=0.22\linewidth]{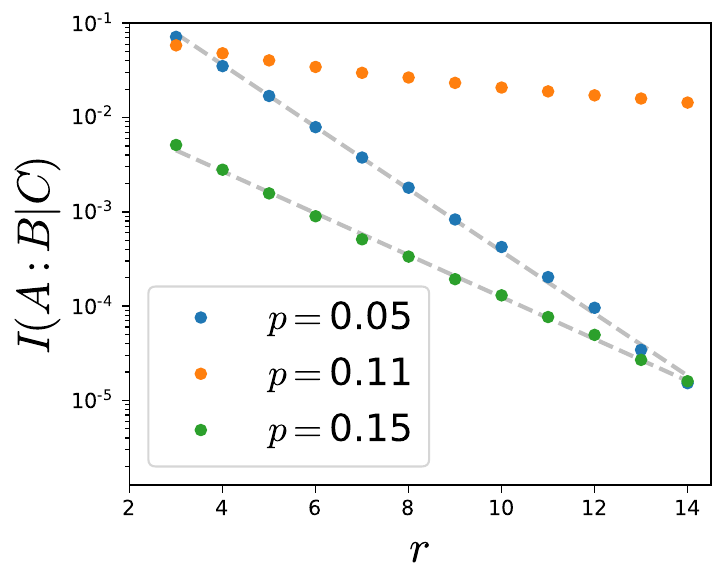}}
    \vspace{-8pt}
    \caption{
    {\bf CMI of dephased toric code state ---}
    \textbf{(a)} Partition with $A$ fixed and varying $r$ (width of $B,C$). 
    \textbf{(b)} $I(A:C|B)$ peaks around $p_c\approx 0.11$ and peak size decays with $r$.
    \textbf{(c)} Finite-size collapse with the scaling ansatz Eq.\eqref{eq: scaling_form}, with $(p_c, \nu, \alpha)=(0.11, 1.8, 1.1)$.
    \textbf{(d)} Above ($p=0.15$) or below  ($p=0.05$) the critical point, CMI decays exponentially with $r$, in contrast to power-law decay at the critical point $p_c \approx 0.11$.
    }
    \label{fig: numerics}
\end{figure*}

\emph{Example: dephased toric code---}
Given that finite $\xi$ implies continuity of a phase, it is natural to expect that a phase transition occurs when $\xi$ diverges. We demonstrate this with a concrete example: toric code topological order subject to dephasing noise. 

Let $\ket{\rm t.c.}$ be a ground state of the toric code Hamiltonian:
\begin{equation}\label{eq: tc_hamiltonian}
    H_{\rm t.c.}= -{\sum}_{\square} A_\square  - {\sum}_{+} B_+,
\end{equation}
where qubits reside on edges of an $L\times L$ square lattice, and $\square$ ($+$) are plaquettes (vertices). 
The two terms are $A_\square=\prod_{i\in\square} X_i$ and $B_+ \equiv \prod_{i\in+}Z_i$ and all mutually commute.  Thus the ground state satisfies all terms, \textit{i.e.} $A_\square\ket{\rm t.c.}=B_+\ket{\rm t.c.}=\ket{\rm t.c.}$.

After applying dephasing noise ${\calE}_{p}[\cdot]\equiv(1-p)(\cdot)+pZ(\cdot)Z$ on each qubit, we obtain a mixed-state  $\rho_p\equiv{\calE}_p^{\otimes L^2}[\ket{\rm t.c.}\bra{\rm t.c.}]$. Physically this corresponds to applying a Pauli-$Z$ gate independently on each qubit with probability $p$. The channel can be realized by evolving the Lindbladian $\calL[\rho]=\sum_i\frac{1}{2}(Z_i\rho Z_i - \rho)$ for time $t_p=-\ln(1-2p)$, with $t=\infty$ corresponding to $p=0.5$.

The mixed state is concisely described by an anyon distribution. We say a plaquette $\square$ is occupied by an anyon if $A_{\square}=-1$ (before decoherence, the ground state has no anyons).  Once a $Z$ operator acts on an edge, it flips the anyon occupancy on the two adjacent plaquettes. Thus for a fixed set of flipped qubits, the resulting state has anyons on plaquettes that contain an odd number of flipped links. The state $\rho_{p}$ is a mixture of all compatible anyon configurations weighted by their probabilities. For a simply connected sub-region $Q$, its reduced density matrix is:
\begin{equation}
\begin{aligned}
    \rho_{p, Q} &= \sum_{\bm_Q} \Pr(\bm_Q) \Pi_{\bm_Q}\\
    \Pr(\bm_Q) &\equiv \sum_{\be} p^{|\be|}(1-p)^{|Q|-|\be|} \delta(\bm_Q=\partial\be)
\end{aligned}
\end{equation}
where the binary vector $\bm_Q$
indicates the anyon configuration of plaquettes within $Q$. 
$\Pi_{\bm_Q}$ is the maximally mixed state 
that has anyon configuration $\bm_Q$ and satisfies $B_+=1$ for all vertices.

After some algebra, one can show that region $Q$'s von Neumann entropy is
\begin{equation}\label{eq: vn_to_shannon}
    S(\rho_{Q, p}) = S(\rho_{Q, 0}) + H(\bm_Q),
\end{equation}
$H(\bm)\equiv-\sum_{\bm}\Pr(\bm)\log\Pr(\bm)$ being the Shannon entropy of the anyon distribution $\Pr(\bm)$. If $Q$ is not simply connected and contains a hole denoted $\Gamma$, the \textit{r.h.s.} of Eq.\eqref{eq: vn_to_shannon} is replaced by $H(\bm_Q, \pi(\bm_\Gamma))$, with $\pi(\bm_\Gamma)$ being the parity of anyon number within $\Gamma$ (see SM for details). Thus for an annulus-shaped $A, B, C$ partition (Fig. \ref{fig: numerics}(a)), 
\begin{align}\label{eq: cmi_into_shannon}
    I(A:C|B) = & H(\bm_{BC},\pi(\bm_A)) - H(\bm_{ABC}) \nonumber\\
    &- H(\bm_{B}, \pi(\bm_{A})) + H(\bm_{AB}).
\end{align}
We simulate $I(A:B|C)$ in this geometry for various $r$s (\textit{i.e.} width of $B$ and $C$) and system sizes by representing $\Pr(\bm)$ as a two-dimensional tensor network and contracting it using the boundary transfer matrix technique~\cite{Murg2007peps, PhysRevA.90.032326} (see SM). 

We first focus on the parameter regime around the presumed critical point $p_c\approx0.11$. Our results (Fig.\ref{fig: main}) show that for any system size $r$, CMI has a peak around $p_c$ with height decreasing with $r$. Furthermore, data from different system sizes can be collapsed with the scaling ansatz: 
\begin{equation}\label{eq: scaling_form}
    I(A:C|B) = r^{-\alpha} \Phi\left( (p-p_c)r^{1/\nu} \right)
\end{equation}
by choosing $p_c=0.11, \alpha=1.1$ and $\nu=1.8$. In particular, when $p=p_c$, CMI decays as a power-law with $r$, in contrast to mutual information and conventional correlation functions. To verify that the state is FML away from the critical point $p_c$, we pick two representative points below and above the threshold: $p=0.05<p_c$ and $p=0.15>p_c$. As shown in Fig.\ref{fig: main}(d), at both points CMI decays exponentially. These observations imply that in the large $r$ limit:
\begin{equation}
    I(A:C|B)\simeq
    \left\{
    \begin{array}{ll}
       e^{-r/\xi_-(p)}  & p<p_c \\
       r^{-\alpha}  & p=p_c\\
       e^{-r/\xi_+(p)}  & p>p_c
    \end{array}
    \right.
\end{equation}
where $\xi_{\pm}(p)$ diverges near the critical point as $\xi_{\pm}\propto(p-p_c)^{-\nu}$. 

Because the Markov length remains finite in $p\in[0, p_c)$ and $p\in(p_c, 0.5]$ \footnote{We remark that the {$p=0.5$} state requires infinite time dephasing Lindbladian evolution. But as we show in SM, a $O(\log L)$ time evolution is enough to obtain a sufficiently close-by state}, these intervals constitute two mixed-state phases. The former is a topologically ordered phase containing $\ket{\rm t.c.}$, and the latter is a trivial phase containing $\rho_{0.5}\propto \sum_{\bs\in{\rm loops}}\ket{\bs}\bra{\bs}$, \textit{i.e.} a classical uniform distribution of all closed-loop spin configurations. This state can be obtained by applying $\calG_{\square}[\cdot]\equiv\frac{1}{2}(\cdot)+\frac{1}{2}A_\square(\cdot)A_\square$ on each plaquette $\square$ of a product state $\ket{\bf 0}\bra{\bf 0}$; thus it belongs to a trivial phase.

Similar to other aspects of decohered toric code studied in~\cite{dennis2002topological, chen2023separability, fan2023diagnostics}, the behavior of CMI can also be understood in terms of the random bond Ising model (RBIM) along the Nishimori line~\cite{ozeki1993phase}. Each term in Eq.\eqref{eq: cmi_into_shannon} can be mapped to a free energy in RBIM (see SM for detailed mapping and RBIM definition). For instance, $H(\bm_{AB}) = \overline{F_{{\rm RBIM}, p}(AB)} + c_1|AB| + c_2$, where $\overline{F_{{\rm RBIM}, p}(AB)}$ is the disorder-averaged free energy of the RBIM defined on region $AB$'s dual lattice,  and $c_{1,2}$ are constants. For non-simply-connected regions $B$ and $BC$, the central hole $A$ is treated as a single dual lattice site in the correponding RBIM. Since $A$ is $O(1)$-sized, from a coarse-grained point of view the CMI is:
\begin{equation}\label{eq: F_defect}
    I(A:C|B) = \overline{F_{\rm def}(4r)} - \overline{F_{\rm def}(2r)},
\end{equation}
where $\overline{F_{\rm def}(x)}$ is the free energy cost of introducing a point defect at the center of RBIM on an $x\times x$ lattice. 
The two terms come from $H(\pi(\bm_A), \bm_{BC})-H(\bm_{ABC})$ and $H(\pi(\bm_A), \bm_{B})-H(\bm_{AB})$, respectively.
The RBIM has ferromagnetic and paramagnetic phases, separated by a critical point presumably described by a conformal field theory, where $\overline{F_{\rm def}(r)}$ has a subleading piece decaying as a power law with $r$. 
Thus we expect the scaling form Eq.\eqref{eq: scaling_form} to originate from the RBIM critical point. As the correlation length is the only length scale near a critical point, we expect that the Markov length $\xi$ can be identified with the RBIM's correlation length. We note however that our fitted exponent $\nu\approx1.8$ deviates from  $\nu_{\rm RBIM}\approx1.5$ reported in~\cite{merz2002rbim}, and we leave this discrepancy, likely due to finite-size effects, for future work.

\emph{$\tilde\calG$ as a quasi-local decoder---}
Using the mapping to RBIM as a bridge, we associate the dephased toric code's mixed-state phase transition to its decodability transition studied in~\cite{dennis2002topological}, since both occur at the RBIM transition. 
In fact in the decodable phase $p<p_c$, the reversal circuit $\tilde\calG$ we constructed works as a quasi-local decoder that recovers quantum information stored in the ground space subspace $V$ of Eq.\eqref{eq: tc_hamiltonian}, \textit{i.e.}
\begin{equation}
    \tilde\calG\circ\calG[\ket{\psi}\bra{\psi}] \approx \ket{\psi}\bra{\psi}\ \ \ \forall \ket{\psi}\in V,
\end{equation}
where $\tilde\calG$ is constructed with respect to an arbitrary initial state in $V$. The claim follows from the observation that $\tilde\calG$ is independent of the choice of $\ket{\psi}$, proved as follows. Each gate $\tilde\calE_{\ell, x}$ depends only on the local density matrix $(\rho_{\ell-1})_{A_{\ell,x}\cup B_{\ell,x}}$ at the $\ell-1$ step. For any observable $Q$ defined on $A_{\ell,x}\cup B_{\ell,x}$, we have $\tr(\rho_{\ell-1} Q) = \bra{\psi}\calC^\dagger_1\calC^\dagger_2...\calC^\dagger_{\ell-1}[Q]\ket{\psi}$. Since $\calC^\dagger_1\calC^\dagger_2...\calC^\dagger_{\ell-1}[Q]$ is a quasi-local operator, its expectation value is independent of $\ket{\psi}$, thus $(\rho_{\ell-1})_{A_{\ell,x}\cup B_{\ell,x}}$ and $\tilde\calE_{\ell, x}$ are also independent of $\ket{\psi}$.

\emph{Discussion---}
We have shown that under a local Lindbladian evolution, a mixed state remains in the same phase of matter whenever its Markov length stays finite. Applying this diagnostic to the dephased toric code state and identifying the Markov length with the RBIM correlation length reveal that the mixed state phase transition and decoding transition precisely coincide and as a byproduct, furnishes a quasi-local decoding channel.  

Our main result and methods can be readily generalized to mixed-state phases with symmetry, \textit{e.g.} symmetry-protected topological (SPT) phases~\cite{de2022symmetry,ma2023average, zhang2022strange, lee2022symmetry, xue2024tensor, guo2024locally, ma2024symmetry}, since the reverse evolution is naturally symmetry respecting: $\tilde{\calG}$ satisfies any strong (or weak) symmetry present in both the forward evolution $\calG$ and the initial state $\rho$. This derives from the same property of the Petz map $\calP(\calE, \rho)$, which is manifest from its definition (see SM). 

Our discussion relies on paths of FML states from local Lindbladian evolution, 
which is only justified if FML states constitute extended regions under generic local Lindbladian evolutions, and non-FML states constitute a measure-zero set requiring fine-tuning (as depicted in Fig.\ref{fig: main}). Our work shows this is true in the case of dephased toric code, but understanding its generic validity is desirable.

The Markov length criteria for continuity of a phase also encompasses pure ground state phases, in which the Markov length should reduce to the correlation length that remains finite along a path of gapped parent Hamiltonians.  It is worth exploring the role of Markov length in understanding phases of other types of mixed-states, \textit{e.g.} Gibbs states and Lindbladian steady states. For the former case, FML property is rigorously established for several special cases~\cite{leifer2008quantum, tomotaka2020clustering, kato2019quantum}, and is believed to be generically true at non-zero temperature. 
On the other hand, there are states with infinite Markov length which are unstable to certain Lindbladian perturbations, \textit{e.g.} the Greenberger–Horne–Zeilinger (GHZ) state $\ket{\rm GHZ}\propto\ket{00...0}+\ket{11...1}$ and critical ground states. It is worth understanding how infinite Markov length is related to the state's instability.

\emph{Acknowledgements---}
We appreciate helpful discussions with Yimu Bao, Matthew P. A. Fisher, Sarang Gopalakrishnan, Tarun Grover, David Huse, Isaac Kim, Tomotaka Kuwahara, Ali Lavasani, Yaodong Li, Ziwen Liu, Ruochen Ma, Bowen Shi and Yijian Zou. We thank Roger G. Melko and Digital Research Alliance of Canada for computational resources.  SS acknowledges the KITP graduate fellow program, during which part of this work was completed. This work was supported by the Perimeter Institute for Theoretical Physics (PI) and the Natural Sciences and Engineering Research Council of Canada (NSERC). Research at PI is supported in part by the Government of Canada through the Department of Innovation, Science and Economic Development Canada and by the Province of Ontario through the Ministry of Colleges and Universities.

\bibliography{main.bib}

\begin{thebibliography}{60}%
\makeatletter
\providecommand \@ifxundefined [1]{%
 \@ifx{#1\undefined}
}%
\providecommand \@ifnum [1]{%
 \ifnum #1\expandafter \@firstoftwo
 \else \expandafter \@secondoftwo
 \fi
}%
\providecommand \@ifx [1]{%
 \ifx #1\expandafter \@firstoftwo
 \else \expandafter \@secondoftwo
 \fi
}%
\providecommand \natexlab [1]{#1}%
\providecommand \enquote  [1]{``#1''}%
\providecommand \bibnamefont  [1]{#1}%
\providecommand \bibfnamefont [1]{#1}%
\providecommand \citenamefont [1]{#1}%
\providecommand \href@noop [0]{\@secondoftwo}%
\providecommand \href [0]{\begingroup \@sanitize@url \@href}%
\providecommand \@href[1]{\@@startlink{#1}\@@href}%
\providecommand \@@href[1]{\endgroup#1\@@endlink}%
\providecommand \@sanitize@url [0]{\catcode `\\12\catcode `\$12\catcode `\&12\catcode `\#12\catcode `\^12\catcode `\_12\catcode `\%12\relax}%
\providecommand \@@startlink[1]{}%
\providecommand \@@endlink[0]{}%
\providecommand \url  [0]{\begingroup\@sanitize@url \@url }%
\providecommand \@url [1]{\endgroup\@href {#1}{\urlprefix }}%
\providecommand \urlprefix  [0]{URL }%
\providecommand \Eprint [0]{\href }%
\providecommand \doibase [0]{https://doi.org/}%
\providecommand \selectlanguage [0]{\@gobble}%
\providecommand \bibinfo  [0]{\@secondoftwo}%
\providecommand \bibfield  [0]{\@secondoftwo}%
\providecommand \translation [1]{[#1]}%
\providecommand \BibitemOpen [0]{}%
\providecommand \bibitemStop [0]{}%
\providecommand \bibitemNoStop [0]{.\EOS\space}%
\providecommand \EOS [0]{\spacefactor3000\relax}%
\providecommand \BibitemShut  [1]{\csname bibitem#1\endcsname}%
\let\auto@bib@innerbib\@empty
\bibitem [{\citenamefont {Verstraete}\ \emph {et~al.}(2009)\citenamefont {Verstraete}, \citenamefont {Wolf},\ and\ \citenamefont {Ignacio~Cirac}}]{diss1}%
  \BibitemOpen
  \bibfield  {author} {\bibinfo {author} {\bibfnamefont {F.}~\bibnamefont {Verstraete}}, \bibinfo {author} {\bibfnamefont {M.~M.}\ \bibnamefont {Wolf}},\ and\ \bibinfo {author} {\bibfnamefont {J.}~\bibnamefont {Ignacio~Cirac}},\ }\bibfield  {title} {\bibinfo {title} {Quantum computation and quantum-state engineering driven by dissipation},\ }\href {https://doi.org/10.1038/nphys1342} {\bibfield  {journal} {\bibinfo  {journal} {Nature Physics}\ }\textbf {\bibinfo {volume} {5}},\ \bibinfo {pages} {633} (\bibinfo {year} {2009})}\BibitemShut {NoStop}%
\bibitem [{\citenamefont {Diehl}\ \emph {et~al.}(2008)\citenamefont {Diehl}, \citenamefont {Micheli}, \citenamefont {Kantian}, \citenamefont {Kraus}, \citenamefont {B{\"u}chler},\ and\ \citenamefont {Zoller}}]{diss2}%
  \BibitemOpen
  \bibfield  {author} {\bibinfo {author} {\bibfnamefont {S.}~\bibnamefont {Diehl}}, \bibinfo {author} {\bibfnamefont {A.}~\bibnamefont {Micheli}}, \bibinfo {author} {\bibfnamefont {A.}~\bibnamefont {Kantian}}, \bibinfo {author} {\bibfnamefont {B.}~\bibnamefont {Kraus}}, \bibinfo {author} {\bibfnamefont {H.~P.}\ \bibnamefont {B{\"u}chler}},\ and\ \bibinfo {author} {\bibfnamefont {P.}~\bibnamefont {Zoller}},\ }\bibfield  {title} {\bibinfo {title} {Quantum states and phases in driven open quantum systems with cold atoms},\ }\href {https://doi.org/10.1038/nphys1073} {\bibfield  {journal} {\bibinfo  {journal} {Nature Physics}\ }\textbf {\bibinfo {volume} {4}},\ \bibinfo {pages} {878} (\bibinfo {year} {2008})}\BibitemShut {NoStop}%
\bibitem [{\citenamefont {Hastings}(2011)}]{hastings2011nonzero}%
  \BibitemOpen
  \bibfield  {author} {\bibinfo {author} {\bibfnamefont {M.~B.}\ \bibnamefont {Hastings}},\ }\bibfield  {title} {\bibinfo {title} {Topological order at nonzero temperature},\ }\href@noop {} {\bibfield  {journal} {\bibinfo  {journal} {Physical review letters}\ }\textbf {\bibinfo {volume} {107}},\ \bibinfo {pages} {210501} (\bibinfo {year} {2011})}\BibitemShut {NoStop}%
\bibitem [{\citenamefont {Coser}\ and\ \citenamefont {P{\'e}rez-Garc{\'\i}a}(2019)}]{coser2019classification}%
  \BibitemOpen
  \bibfield  {author} {\bibinfo {author} {\bibfnamefont {A.}~\bibnamefont {Coser}}\ and\ \bibinfo {author} {\bibfnamefont {D.}~\bibnamefont {P{\'e}rez-Garc{\'\i}a}},\ }\bibfield  {title} {\bibinfo {title} {Classification of phases for mixed states via fast dissipative evolution},\ }\href@noop {} {\bibfield  {journal} {\bibinfo  {journal} {Quantum}\ }\textbf {\bibinfo {volume} {3}},\ \bibinfo {pages} {174} (\bibinfo {year} {2019})}\BibitemShut {NoStop}%
\bibitem [{\citenamefont {Fan}\ \emph {et~al.}(2023)\citenamefont {Fan}, \citenamefont {Bao}, \citenamefont {Altman},\ and\ \citenamefont {Vishwanath}}]{fan2023diagnostics}%
  \BibitemOpen
  \bibfield  {author} {\bibinfo {author} {\bibfnamefont {R.}~\bibnamefont {Fan}}, \bibinfo {author} {\bibfnamefont {Y.}~\bibnamefont {Bao}}, \bibinfo {author} {\bibfnamefont {E.}~\bibnamefont {Altman}},\ and\ \bibinfo {author} {\bibfnamefont {A.}~\bibnamefont {Vishwanath}},\ }\bibfield  {title} {\bibinfo {title} {Diagnostics of mixed-state topological order and breakdown of quantum memory},\ }\href@noop {} {\bibfield  {journal} {\bibinfo  {journal} {arXiv preprint arXiv:2301.05689}\ } (\bibinfo {year} {2023})}\BibitemShut {NoStop}%
\bibitem [{\citenamefont {Bao}\ \emph {et~al.}(2023)\citenamefont {Bao}, \citenamefont {Fan}, \citenamefont {Vishwanath},\ and\ \citenamefont {Altman}}]{bao2023mixed}%
  \BibitemOpen
  \bibfield  {author} {\bibinfo {author} {\bibfnamefont {Y.}~\bibnamefont {Bao}}, \bibinfo {author} {\bibfnamefont {R.}~\bibnamefont {Fan}}, \bibinfo {author} {\bibfnamefont {A.}~\bibnamefont {Vishwanath}},\ and\ \bibinfo {author} {\bibfnamefont {E.}~\bibnamefont {Altman}},\ }\bibfield  {title} {\bibinfo {title} {Mixed-state topological order and the errorfield double formulation of decoherence-induced transitions},\ }\href@noop {} {\bibfield  {journal} {\bibinfo  {journal} {arXiv preprint arXiv:2301.05687}\ } (\bibinfo {year} {2023})}\BibitemShut {NoStop}%
\bibitem [{\citenamefont {Lee}\ \emph {et~al.}(2023)\citenamefont {Lee}, \citenamefont {Jian},\ and\ \citenamefont {Xu}}]{lee2023quantum}%
  \BibitemOpen
  \bibfield  {author} {\bibinfo {author} {\bibfnamefont {J.~Y.}\ \bibnamefont {Lee}}, \bibinfo {author} {\bibfnamefont {C.-M.}\ \bibnamefont {Jian}},\ and\ \bibinfo {author} {\bibfnamefont {C.}~\bibnamefont {Xu}},\ }\bibfield  {title} {\bibinfo {title} {Quantum criticality under decoherence or weak measurement},\ }\href@noop {} {\bibfield  {journal} {\bibinfo  {journal} {arXiv preprint arXiv:2301.05238}\ } (\bibinfo {year} {2023})}\BibitemShut {NoStop}%
\bibitem [{\citenamefont {Zou}\ \emph {et~al.}(2023)\citenamefont {Zou}, \citenamefont {Sang},\ and\ \citenamefont {Hsieh}}]{zou2023channeling}%
  \BibitemOpen
  \bibfield  {author} {\bibinfo {author} {\bibfnamefont {Y.}~\bibnamefont {Zou}}, \bibinfo {author} {\bibfnamefont {S.}~\bibnamefont {Sang}},\ and\ \bibinfo {author} {\bibfnamefont {T.~H.}\ \bibnamefont {Hsieh}},\ }\bibfield  {title} {\bibinfo {title} {Channeling quantum criticality},\ }\href@noop {} {\bibfield  {journal} {\bibinfo  {journal} {Physical Review Letters}\ }\textbf {\bibinfo {volume} {130}},\ \bibinfo {pages} {250403} (\bibinfo {year} {2023})}\BibitemShut {NoStop}%
\bibitem [{\citenamefont {de~Groot}\ \emph {et~al.}(2022)\citenamefont {de~Groot}, \citenamefont {Turzillo},\ and\ \citenamefont {Schuch}}]{de2022symmetry}%
  \BibitemOpen
  \bibfield  {author} {\bibinfo {author} {\bibfnamefont {C.}~\bibnamefont {de~Groot}}, \bibinfo {author} {\bibfnamefont {A.}~\bibnamefont {Turzillo}},\ and\ \bibinfo {author} {\bibfnamefont {N.}~\bibnamefont {Schuch}},\ }\bibfield  {title} {\bibinfo {title} {Symmetry protected topological order in open quantum systems},\ }\href@noop {} {\bibfield  {journal} {\bibinfo  {journal} {Quantum}\ }\textbf {\bibinfo {volume} {6}},\ \bibinfo {pages} {856} (\bibinfo {year} {2022})}\BibitemShut {NoStop}%
\bibitem [{\citenamefont {Ma}\ and\ \citenamefont {Wang}(2023)}]{ma2023average}%
  \BibitemOpen
  \bibfield  {author} {\bibinfo {author} {\bibfnamefont {R.}~\bibnamefont {Ma}}\ and\ \bibinfo {author} {\bibfnamefont {C.}~\bibnamefont {Wang}},\ }\bibfield  {title} {\bibinfo {title} {Average symmetry-protected topological phases},\ }\href@noop {} {\bibfield  {journal} {\bibinfo  {journal} {Physical Review X}\ }\textbf {\bibinfo {volume} {13}},\ \bibinfo {pages} {031016} (\bibinfo {year} {2023})}\BibitemShut {NoStop}%
\bibitem [{\citenamefont {Zhang}\ \emph {et~al.}(2022)\citenamefont {Zhang}, \citenamefont {Qi},\ and\ \citenamefont {Bi}}]{zhang2022strange}%
  \BibitemOpen
  \bibfield  {author} {\bibinfo {author} {\bibfnamefont {J.-H.}\ \bibnamefont {Zhang}}, \bibinfo {author} {\bibfnamefont {Y.}~\bibnamefont {Qi}},\ and\ \bibinfo {author} {\bibfnamefont {Z.}~\bibnamefont {Bi}},\ }\href@noop {} {\bibinfo {title} {Strange correlation function for average symmetry-protected topological phases}} (\bibinfo {year} {2022}),\ \Eprint {https://arxiv.org/abs/2210.17485} {arXiv:2210.17485 [cond-mat.str-el]} \BibitemShut {NoStop}%
\bibitem [{\citenamefont {Ma}\ \emph {et~al.}(2023)\citenamefont {Ma}, \citenamefont {Zhang}, \citenamefont {Bi}, \citenamefont {Cheng},\ and\ \citenamefont {Wang}}]{ma2023topological}%
  \BibitemOpen
  \bibfield  {author} {\bibinfo {author} {\bibfnamefont {R.}~\bibnamefont {Ma}}, \bibinfo {author} {\bibfnamefont {J.-H.}\ \bibnamefont {Zhang}}, \bibinfo {author} {\bibfnamefont {Z.}~\bibnamefont {Bi}}, \bibinfo {author} {\bibfnamefont {M.}~\bibnamefont {Cheng}},\ and\ \bibinfo {author} {\bibfnamefont {C.}~\bibnamefont {Wang}},\ }\href@noop {} {\bibinfo {title} {Topological phases with average symmetries: the decohered, the disordered, and the intrinsic}} (\bibinfo {year} {2023}),\ \Eprint {https://arxiv.org/abs/2305.16399} {arXiv:2305.16399 [cond-mat.str-el]} \BibitemShut {NoStop}%
\bibitem [{\citenamefont {Rakovszky}\ \emph {et~al.}(2024)\citenamefont {Rakovszky}, \citenamefont {Gopalakrishnan},\ and\ \citenamefont {von Keyserlingk}}]{rakovszky2024defining}%
  \BibitemOpen
  \bibfield  {author} {\bibinfo {author} {\bibfnamefont {T.}~\bibnamefont {Rakovszky}}, \bibinfo {author} {\bibfnamefont {S.}~\bibnamefont {Gopalakrishnan}},\ and\ \bibinfo {author} {\bibfnamefont {C.}~\bibnamefont {von Keyserlingk}},\ }\href@noop {} {\bibinfo {title} {Defining stable phases of open quantum systems}} (\bibinfo {year} {2024}),\ \Eprint {https://arxiv.org/abs/2308.15495} {arXiv:2308.15495 [quant-ph]} \BibitemShut {NoStop}%
\bibitem [{\citenamefont {Sang}\ \emph {et~al.}(2023{\natexlab{a}})\citenamefont {Sang}, \citenamefont {Zou},\ and\ \citenamefont {Hsieh}}]{sang2023mixed}%
  \BibitemOpen
  \bibfield  {author} {\bibinfo {author} {\bibfnamefont {S.}~\bibnamefont {Sang}}, \bibinfo {author} {\bibfnamefont {Y.}~\bibnamefont {Zou}},\ and\ \bibinfo {author} {\bibfnamefont {T.~H.}\ \bibnamefont {Hsieh}},\ }\bibfield  {title} {\bibinfo {title} {Mixed-state quantum phases: Renormalization and quantum error correction},\ }\href@noop {} {\bibfield  {journal} {\bibinfo  {journal} {arXiv preprint arXiv:2310.08639}\ } (\bibinfo {year} {2023}{\natexlab{a}})}\BibitemShut {NoStop}%
\bibitem [{\citenamefont {Lu}\ \emph {et~al.}(2023)\citenamefont {Lu}, \citenamefont {Zhang}, \citenamefont {Vijay},\ and\ \citenamefont {Hsieh}}]{lu2023mixed}%
  \BibitemOpen
  \bibfield  {author} {\bibinfo {author} {\bibfnamefont {T.-C.}\ \bibnamefont {Lu}}, \bibinfo {author} {\bibfnamefont {Z.}~\bibnamefont {Zhang}}, \bibinfo {author} {\bibfnamefont {S.}~\bibnamefont {Vijay}},\ and\ \bibinfo {author} {\bibfnamefont {T.~H.}\ \bibnamefont {Hsieh}},\ }\bibfield  {title} {\bibinfo {title} {Mixed-state long-range order and criticality from measurement and feedback},\ }\href {https://doi.org/10.1103/PRXQuantum.4.030318} {\bibfield  {journal} {\bibinfo  {journal} {PRX Quantum}\ }\textbf {\bibinfo {volume} {4}},\ \bibinfo {pages} {030318} (\bibinfo {year} {2023})}\BibitemShut {NoStop}%
\bibitem [{\citenamefont {Lee}\ \emph {et~al.}(2022{\natexlab{a}})\citenamefont {Lee}, \citenamefont {Ji}, \citenamefont {Bi},\ and\ \citenamefont {Fisher}}]{lee2022decoding}%
  \BibitemOpen
  \bibfield  {author} {\bibinfo {author} {\bibfnamefont {J.~Y.}\ \bibnamefont {Lee}}, \bibinfo {author} {\bibfnamefont {W.}~\bibnamefont {Ji}}, \bibinfo {author} {\bibfnamefont {Z.}~\bibnamefont {Bi}},\ and\ \bibinfo {author} {\bibfnamefont {M.~P.~A.}\ \bibnamefont {Fisher}},\ }\href@noop {} {\bibinfo {title} {Decoding measurement-prepared quantum phases and transitions: from ising model to gauge theory, and beyond}} (\bibinfo {year} {2022}{\natexlab{a}}),\ \Eprint {https://arxiv.org/abs/2208.11699} {arXiv:2208.11699 [cond-mat.str-el]} \BibitemShut {NoStop}%
\bibitem [{\citenamefont {Zhu}\ \emph {et~al.}(2022)\citenamefont {Zhu}, \citenamefont {Tantivasadakarn}, \citenamefont {Vishwanath}, \citenamefont {Trebst},\ and\ \citenamefont {Verresen}}]{zhu2022nishimoris}%
  \BibitemOpen
  \bibfield  {author} {\bibinfo {author} {\bibfnamefont {G.-Y.}\ \bibnamefont {Zhu}}, \bibinfo {author} {\bibfnamefont {N.}~\bibnamefont {Tantivasadakarn}}, \bibinfo {author} {\bibfnamefont {A.}~\bibnamefont {Vishwanath}}, \bibinfo {author} {\bibfnamefont {S.}~\bibnamefont {Trebst}},\ and\ \bibinfo {author} {\bibfnamefont {R.}~\bibnamefont {Verresen}},\ }\href@noop {} {\bibinfo {title} {Nishimori's cat: stable long-range entanglement from finite-depth unitaries and weak measurements}} (\bibinfo {year} {2022}),\ \Eprint {https://arxiv.org/abs/2208.11136} {arXiv:2208.11136 [quant-ph]} \BibitemShut {NoStop}%
\bibitem [{\citenamefont {Chen}\ and\ \citenamefont {Grover}(2023{\natexlab{a}})}]{chen2023symmetryenforced}%
  \BibitemOpen
  \bibfield  {author} {\bibinfo {author} {\bibfnamefont {Y.-H.}\ \bibnamefont {Chen}}\ and\ \bibinfo {author} {\bibfnamefont {T.}~\bibnamefont {Grover}},\ }\href@noop {} {\bibinfo {title} {Symmetry-enforced many-body separability transitions}} (\bibinfo {year} {2023}{\natexlab{a}}),\ \Eprint {https://arxiv.org/abs/2310.07286} {arXiv:2310.07286 [quant-ph]} \BibitemShut {NoStop}%
\bibitem [{\citenamefont {Chen}\ and\ \citenamefont {Grover}(2023{\natexlab{b}})}]{chen2023separability}%
  \BibitemOpen
  \bibfield  {author} {\bibinfo {author} {\bibfnamefont {Y.-H.}\ \bibnamefont {Chen}}\ and\ \bibinfo {author} {\bibfnamefont {T.}~\bibnamefont {Grover}},\ }\bibfield  {title} {\bibinfo {title} {Separability transitions in topological states induced by local decoherence},\ }\href@noop {} {\bibfield  {journal} {\bibinfo  {journal} {arXiv preprint arXiv:2309.11879}\ } (\bibinfo {year} {2023}{\natexlab{b}})}\BibitemShut {NoStop}%
\bibitem [{\citenamefont {Lessa}\ \emph {et~al.}(2024)\citenamefont {Lessa}, \citenamefont {Cheng},\ and\ \citenamefont {Wang}}]{lessa2024mixedstate}%
  \BibitemOpen
  \bibfield  {author} {\bibinfo {author} {\bibfnamefont {L.~A.}\ \bibnamefont {Lessa}}, \bibinfo {author} {\bibfnamefont {M.}~\bibnamefont {Cheng}},\ and\ \bibinfo {author} {\bibfnamefont {C.}~\bibnamefont {Wang}},\ }\href@noop {} {\bibinfo {title} {Mixed-state quantum anomaly and multipartite entanglement}} (\bibinfo {year} {2024}),\ \Eprint {https://arxiv.org/abs/2401.17357} {arXiv:2401.17357 [cond-mat.str-el]} \BibitemShut {NoStop}%
\bibitem [{\citenamefont {Chen}\ and\ \citenamefont {Grover}(2024)}]{chen2024unconventional}%
  \BibitemOpen
  \bibfield  {author} {\bibinfo {author} {\bibfnamefont {Y.-H.}\ \bibnamefont {Chen}}\ and\ \bibinfo {author} {\bibfnamefont {T.}~\bibnamefont {Grover}},\ }\href@noop {} {\bibinfo {title} {Unconventional topological mixed-state transition and critical phase induced by self-dual coherent errors}} (\bibinfo {year} {2024}),\ \Eprint {https://arxiv.org/abs/2403.06553} {arXiv:2403.06553 [quant-ph]} \BibitemShut {NoStop}%
\bibitem [{\citenamefont {Lee}\ \emph {et~al.}(2022{\natexlab{b}})\citenamefont {Lee}, \citenamefont {You},\ and\ \citenamefont {Xu}}]{lee2022symmetry}%
  \BibitemOpen
  \bibfield  {author} {\bibinfo {author} {\bibfnamefont {J.~Y.}\ \bibnamefont {Lee}}, \bibinfo {author} {\bibfnamefont {Y.-Z.}\ \bibnamefont {You}},\ and\ \bibinfo {author} {\bibfnamefont {C.}~\bibnamefont {Xu}},\ }\bibfield  {title} {\bibinfo {title} {Symmetry protected topological phases under decoherence},\ }\href@noop {} {\bibfield  {journal} {\bibinfo  {journal} {arXiv preprint arXiv:2210.16323}\ } (\bibinfo {year} {2022}{\natexlab{b}})}\BibitemShut {NoStop}%
\bibitem [{\citenamefont {Lee}(2024)}]{lee2024exact}%
  \BibitemOpen
  \bibfield  {author} {\bibinfo {author} {\bibfnamefont {J.~Y.}\ \bibnamefont {Lee}},\ }\href@noop {} {\bibinfo {title} {Exact calculations of coherent information for toric codes under decoherence: Identifying the fundamental error threshold}} (\bibinfo {year} {2024}),\ \Eprint {https://arxiv.org/abs/2402.16937} {arXiv:2402.16937 [cond-mat.stat-mech]} \BibitemShut {NoStop}%
\bibitem [{\citenamefont {Sohal}\ and\ \citenamefont {Prem}(2024)}]{sohal2024noisy}%
  \BibitemOpen
  \bibfield  {author} {\bibinfo {author} {\bibfnamefont {R.}~\bibnamefont {Sohal}}\ and\ \bibinfo {author} {\bibfnamefont {A.}~\bibnamefont {Prem}},\ }\href@noop {} {\bibinfo {title} {A noisy approach to intrinsically mixed-state topological order}} (\bibinfo {year} {2024}),\ \Eprint {https://arxiv.org/abs/2403.13879} {arXiv:2403.13879 [cond-mat.str-el]} \BibitemShut {NoStop}%
\bibitem [{\citenamefont {Ellison}\ and\ \citenamefont {Cheng}(2024)}]{ellison2024towards}%
  \BibitemOpen
  \bibfield  {author} {\bibinfo {author} {\bibfnamefont {T.}~\bibnamefont {Ellison}}\ and\ \bibinfo {author} {\bibfnamefont {M.}~\bibnamefont {Cheng}},\ }\bibfield  {title} {\bibinfo {title} {Towards a classification of mixed-state topological orders in two dimensions},\ }\href@noop {} {\bibfield  {journal} {\bibinfo  {journal} {arXiv preprint arXiv:2405.02390}\ } (\bibinfo {year} {2024})}\BibitemShut {NoStop}%
\bibitem [{\citenamefont {Wang}\ \emph {et~al.}(2023{\natexlab{a}})\citenamefont {Wang}, \citenamefont {Wu},\ and\ \citenamefont {Wang}}]{wang2023intrinsic}%
  \BibitemOpen
  \bibfield  {author} {\bibinfo {author} {\bibfnamefont {Z.}~\bibnamefont {Wang}}, \bibinfo {author} {\bibfnamefont {Z.}~\bibnamefont {Wu}},\ and\ \bibinfo {author} {\bibfnamefont {Z.}~\bibnamefont {Wang}},\ }\bibfield  {title} {\bibinfo {title} {Intrinsic mixed-state topological order without quantum memory},\ }\href@noop {} {\bibfield  {journal} {\bibinfo  {journal} {arXiv preprint arXiv:2307.13758}\ } (\bibinfo {year} {2023}{\natexlab{a}})}\BibitemShut {NoStop}%
\bibitem [{\citenamefont {Wang}\ \emph {et~al.}(2023{\natexlab{b}})\citenamefont {Wang}, \citenamefont {Dai}, \citenamefont {Wang},\ and\ \citenamefont {Wang}}]{wang2023topologically}%
  \BibitemOpen
  \bibfield  {author} {\bibinfo {author} {\bibfnamefont {Z.}~\bibnamefont {Wang}}, \bibinfo {author} {\bibfnamefont {X.-D.}\ \bibnamefont {Dai}}, \bibinfo {author} {\bibfnamefont {H.-R.}\ \bibnamefont {Wang}},\ and\ \bibinfo {author} {\bibfnamefont {Z.}~\bibnamefont {Wang}},\ }\bibfield  {title} {\bibinfo {title} {Topologically ordered steady states in open quantum systems},\ }\href@noop {} {\bibfield  {journal} {\bibinfo  {journal} {arXiv preprint arXiv:2306.12482}\ } (\bibinfo {year} {2023}{\natexlab{b}})}\BibitemShut {NoStop}%
\bibitem [{\citenamefont {Wang}\ and\ \citenamefont {Li}(2024)}]{wang2024anomaly}%
  \BibitemOpen
  \bibfield  {author} {\bibinfo {author} {\bibfnamefont {Z.}~\bibnamefont {Wang}}\ and\ \bibinfo {author} {\bibfnamefont {L.}~\bibnamefont {Li}},\ }\bibfield  {title} {\bibinfo {title} {Anomaly in open quantum systems and its implications on mixed-state quantum phases},\ }\href@noop {} {\bibfield  {journal} {\bibinfo  {journal} {arXiv preprint arXiv:2403.14533}\ } (\bibinfo {year} {2024})}\BibitemShut {NoStop}%
\bibitem [{\citenamefont {Xue}\ \emph {et~al.}(2024)\citenamefont {Xue}, \citenamefont {Lee},\ and\ \citenamefont {Bao}}]{xue2024tensor}%
  \BibitemOpen
  \bibfield  {author} {\bibinfo {author} {\bibfnamefont {H.}~\bibnamefont {Xue}}, \bibinfo {author} {\bibfnamefont {J.~Y.}\ \bibnamefont {Lee}},\ and\ \bibinfo {author} {\bibfnamefont {Y.}~\bibnamefont {Bao}},\ }\bibfield  {title} {\bibinfo {title} {Tensor network formulation of symmetry protected topological phases in mixed states},\ }\href@noop {} {\bibfield  {journal} {\bibinfo  {journal} {arXiv preprint arXiv:2403.17069}\ } (\bibinfo {year} {2024})}\BibitemShut {NoStop}%
\bibitem [{\citenamefont {Guo}\ \emph {et~al.}(2024)\citenamefont {Guo}, \citenamefont {Zhang}, \citenamefont {Yang},\ and\ \citenamefont {Bi}}]{guo2024locally}%
  \BibitemOpen
  \bibfield  {author} {\bibinfo {author} {\bibfnamefont {Y.}~\bibnamefont {Guo}}, \bibinfo {author} {\bibfnamefont {J.-H.}\ \bibnamefont {Zhang}}, \bibinfo {author} {\bibfnamefont {S.}~\bibnamefont {Yang}},\ and\ \bibinfo {author} {\bibfnamefont {Z.}~\bibnamefont {Bi}},\ }\bibfield  {title} {\bibinfo {title} {Locally purified density operators for symmetry-protected topological phases in mixed states},\ }\href@noop {} {\bibfield  {journal} {\bibinfo  {journal} {arXiv preprint arXiv:2403.16978}\ } (\bibinfo {year} {2024})}\BibitemShut {NoStop}%
\bibitem [{\citenamefont {Ma}\ and\ \citenamefont {Turzillo}(2024)}]{ma2024symmetry}%
  \BibitemOpen
  \bibfield  {author} {\bibinfo {author} {\bibfnamefont {R.}~\bibnamefont {Ma}}\ and\ \bibinfo {author} {\bibfnamefont {A.}~\bibnamefont {Turzillo}},\ }\bibfield  {title} {\bibinfo {title} {Symmetry protected topological phases of mixed states in the doubled space},\ }\href@noop {} {\bibfield  {journal} {\bibinfo  {journal} {arXiv preprint arXiv:2403.13280}\ } (\bibinfo {year} {2024})}\BibitemShut {NoStop}%
\bibitem [{\citenamefont {Li}\ and\ \citenamefont {Fisher}(2021)}]{li2021robust}%
  \BibitemOpen
  \bibfield  {author} {\bibinfo {author} {\bibfnamefont {Y.}~\bibnamefont {Li}}\ and\ \bibinfo {author} {\bibfnamefont {M.}~\bibnamefont {Fisher}},\ }\bibfield  {title} {\bibinfo {title} {Robust decoding in monitored dynamics of open quantum systems with z\_2 symmetry},\ }\href@noop {} {\bibfield  {journal} {\bibinfo  {journal} {arXiv preprint arXiv:2108.04274}\ } (\bibinfo {year} {2021})}\BibitemShut {NoStop}%
\bibitem [{\citenamefont {Hastings}\ and\ \citenamefont {Koma}(2006)}]{hastings2006spectral}%
  \BibitemOpen
  \bibfield  {author} {\bibinfo {author} {\bibfnamefont {M.~B.}\ \bibnamefont {Hastings}}\ and\ \bibinfo {author} {\bibfnamefont {T.}~\bibnamefont {Koma}},\ }\bibfield  {title} {\bibinfo {title} {Spectral gap and exponential decay of correlations},\ }\href@noop {} {\bibfield  {journal} {\bibinfo  {journal} {Communications in mathematical physics}\ }\textbf {\bibinfo {volume} {265}},\ \bibinfo {pages} {781} (\bibinfo {year} {2006})}\BibitemShut {NoStop}%
\bibitem [{\citenamefont {Bravyi}\ \emph {et~al.}(2010)\citenamefont {Bravyi}, \citenamefont {Hastings},\ and\ \citenamefont {Michalakis}}]{bravyi2010topological}%
  \BibitemOpen
  \bibfield  {author} {\bibinfo {author} {\bibfnamefont {S.}~\bibnamefont {Bravyi}}, \bibinfo {author} {\bibfnamefont {M.~B.}\ \bibnamefont {Hastings}},\ and\ \bibinfo {author} {\bibfnamefont {S.}~\bibnamefont {Michalakis}},\ }\bibfield  {title} {\bibinfo {title} {Topological quantum order: stability under local perturbations},\ }\href@noop {} {\bibfield  {journal} {\bibinfo  {journal} {Journal of mathematical physics}\ }\textbf {\bibinfo {volume} {51}} (\bibinfo {year} {2010})}\BibitemShut {NoStop}%
\bibitem [{\citenamefont {Michalakis}\ and\ \citenamefont {Zwolak}(2013)}]{michalakis2013stability}%
  \BibitemOpen
  \bibfield  {author} {\bibinfo {author} {\bibfnamefont {S.}~\bibnamefont {Michalakis}}\ and\ \bibinfo {author} {\bibfnamefont {J.~P.}\ \bibnamefont {Zwolak}},\ }\bibfield  {title} {\bibinfo {title} {Stability of frustration-free hamiltonians},\ }\href@noop {} {\bibfield  {journal} {\bibinfo  {journal} {Communications in Mathematical Physics}\ }\textbf {\bibinfo {volume} {322}},\ \bibinfo {pages} {277} (\bibinfo {year} {2013})}\BibitemShut {NoStop}%
\bibitem [{\citenamefont {Cubitt}\ \emph {et~al.}(2015)\citenamefont {Cubitt}, \citenamefont {Lucia}, \citenamefont {Michalakis},\ and\ \citenamefont {Perez-Garcia}}]{cubitt2015stability}%
  \BibitemOpen
  \bibfield  {author} {\bibinfo {author} {\bibfnamefont {T.~S.}\ \bibnamefont {Cubitt}}, \bibinfo {author} {\bibfnamefont {A.}~\bibnamefont {Lucia}}, \bibinfo {author} {\bibfnamefont {S.}~\bibnamefont {Michalakis}},\ and\ \bibinfo {author} {\bibfnamefont {D.}~\bibnamefont {Perez-Garcia}},\ }\bibfield  {title} {\bibinfo {title} {Stability of local quantum dissipative systems},\ }\href@noop {} {\bibfield  {journal} {\bibinfo  {journal} {Communications in Mathematical Physics}\ }\textbf {\bibinfo {volume} {337}},\ \bibinfo {pages} {1275} (\bibinfo {year} {2015})}\BibitemShut {NoStop}%
\bibitem [{\citenamefont {Liu}\ and\ \citenamefont {Lieu}(2024)}]{liu2024dissipative}%
  \BibitemOpen
  \bibfield  {author} {\bibinfo {author} {\bibfnamefont {Y.-J.}\ \bibnamefont {Liu}}\ and\ \bibinfo {author} {\bibfnamefont {S.}~\bibnamefont {Lieu}},\ }\bibfield  {title} {\bibinfo {title} {Dissipative phase transitions and passive error correction},\ }\href@noop {} {\bibfield  {journal} {\bibinfo  {journal} {Physical Review A}\ }\textbf {\bibinfo {volume} {109}},\ \bibinfo {pages} {022422} (\bibinfo {year} {2024})}\BibitemShut {NoStop}%
\bibitem [{\citenamefont {Kitaev}\ and\ \citenamefont {Preskill}(2006)}]{PhysRevLett.96.110404}%
  \BibitemOpen
  \bibfield  {author} {\bibinfo {author} {\bibfnamefont {A.}~\bibnamefont {Kitaev}}\ and\ \bibinfo {author} {\bibfnamefont {J.}~\bibnamefont {Preskill}},\ }\bibfield  {title} {\bibinfo {title} {Topological entanglement entropy},\ }\href {https://doi.org/10.1103/PhysRevLett.96.110404} {\bibfield  {journal} {\bibinfo  {journal} {Phys. Rev. Lett.}\ }\textbf {\bibinfo {volume} {96}},\ \bibinfo {pages} {110404} (\bibinfo {year} {2006})}\BibitemShut {NoStop}%
\bibitem [{\citenamefont {Levin}\ and\ \citenamefont {Wen}(2006)}]{PhysRevLett.96.110405}%
  \BibitemOpen
  \bibfield  {author} {\bibinfo {author} {\bibfnamefont {M.}~\bibnamefont {Levin}}\ and\ \bibinfo {author} {\bibfnamefont {X.-G.}\ \bibnamefont {Wen}},\ }\bibfield  {title} {\bibinfo {title} {Detecting topological order in a ground state wave function},\ }\href {https://doi.org/10.1103/PhysRevLett.96.110405} {\bibfield  {journal} {\bibinfo  {journal} {Phys. Rev. Lett.}\ }\textbf {\bibinfo {volume} {96}},\ \bibinfo {pages} {110405} (\bibinfo {year} {2006})}\BibitemShut {NoStop}%
\bibitem [{\citenamefont {Shi}\ \emph {et~al.}(2020)\citenamefont {Shi}, \citenamefont {Kato},\ and\ \citenamefont {Kim}}]{SHI2020168164}%
  \BibitemOpen
  \bibfield  {author} {\bibinfo {author} {\bibfnamefont {B.}~\bibnamefont {Shi}}, \bibinfo {author} {\bibfnamefont {K.}~\bibnamefont {Kato}},\ and\ \bibinfo {author} {\bibfnamefont {I.~H.}\ \bibnamefont {Kim}},\ }\bibfield  {title} {\bibinfo {title} {Fusion rules from entanglement},\ }\href {https://doi.org/https://doi.org/10.1016/j.aop.2020.168164} {\bibfield  {journal} {\bibinfo  {journal} {Annals of Physics}\ }\textbf {\bibinfo {volume} {418}},\ \bibinfo {pages} {168164} (\bibinfo {year} {2020})}\BibitemShut {NoStop}%
\bibitem [{\citenamefont {Sang}\ \emph {et~al.}(2023{\natexlab{b}})\citenamefont {Sang}, \citenamefont {Li}, \citenamefont {Hsieh},\ and\ \citenamefont {Yoshida}}]{sanghybrid}%
  \BibitemOpen
  \bibfield  {author} {\bibinfo {author} {\bibfnamefont {S.}~\bibnamefont {Sang}}, \bibinfo {author} {\bibfnamefont {Z.}~\bibnamefont {Li}}, \bibinfo {author} {\bibfnamefont {T.~H.}\ \bibnamefont {Hsieh}},\ and\ \bibinfo {author} {\bibfnamefont {B.}~\bibnamefont {Yoshida}},\ }\bibfield  {title} {\bibinfo {title} {Ultrafast entanglement dynamics in monitored quantum circuits},\ }\href {https://doi.org/10.1103/PRXQuantum.4.040332} {\bibfield  {journal} {\bibinfo  {journal} {PRX Quantum}\ }\textbf {\bibinfo {volume} {4}},\ \bibinfo {pages} {040332} (\bibinfo {year} {2023}{\natexlab{b}})}\BibitemShut {NoStop}%
\bibitem [{\citenamefont {Zhang}\ and\ \citenamefont {Gopalakrishnan}(2024)}]{zhang2024nonlocal}%
  \BibitemOpen
  \bibfield  {author} {\bibinfo {author} {\bibfnamefont {Y.}~\bibnamefont {Zhang}}\ and\ \bibinfo {author} {\bibfnamefont {S.}~\bibnamefont {Gopalakrishnan}},\ }\href@noop {} {\bibinfo {title} {Nonlocal growth of quantum conditional mutual information under decoherence}} (\bibinfo {year} {2024}),\ \Eprint {https://arxiv.org/abs/2402.03439} {arXiv:2402.03439 [quant-ph]} \BibitemShut {NoStop}%
\bibitem [{\citenamefont {un~Lee}\ \emph {et~al.}(2024)\citenamefont {un~Lee}, \citenamefont {Oh}, \citenamefont {Wong}, \citenamefont {Chen},\ and\ \citenamefont {Jiang}}]{lee2024universal}%
  \BibitemOpen
  \bibfield  {author} {\bibinfo {author} {\bibfnamefont {S.}~\bibnamefont {un~Lee}}, \bibinfo {author} {\bibfnamefont {C.}~\bibnamefont {Oh}}, \bibinfo {author} {\bibfnamefont {Y.}~\bibnamefont {Wong}}, \bibinfo {author} {\bibfnamefont {S.}~\bibnamefont {Chen}},\ and\ \bibinfo {author} {\bibfnamefont {L.}~\bibnamefont {Jiang}},\ }\href@noop {} {\bibinfo {title} {Universal spreading of conditional mutual information in noisy random circuits}} (\bibinfo {year} {2024}),\ \Eprint {https://arxiv.org/abs/2402.18548} {arXiv:2402.18548 [quant-ph]} \BibitemShut {NoStop}%
\bibitem [{\citenamefont {Onorati}\ \emph {et~al.}(2023)\citenamefont {Onorati}, \citenamefont {Rouz{\'e}}, \citenamefont {Fran{\c{c}}a},\ and\ \citenamefont {Watson}}]{onorati2023efficient}%
  \BibitemOpen
  \bibfield  {author} {\bibinfo {author} {\bibfnamefont {E.}~\bibnamefont {Onorati}}, \bibinfo {author} {\bibfnamefont {C.}~\bibnamefont {Rouz{\'e}}}, \bibinfo {author} {\bibfnamefont {D.~S.}\ \bibnamefont {Fran{\c{c}}a}},\ and\ \bibinfo {author} {\bibfnamefont {J.~D.}\ \bibnamefont {Watson}},\ }\bibfield  {title} {\bibinfo {title} {Efficient learning of ground \& thermal states within phases of matter},\ }\href@noop {} {\bibfield  {journal} {\bibinfo  {journal} {arXiv preprint arXiv:2301.12946}\ } (\bibinfo {year} {2023})}\BibitemShut {NoStop}%
\bibitem [{\citenamefont {Brandao}\ and\ \citenamefont {Kastoryano}(2019)}]{brandao2019finite}%
  \BibitemOpen
  \bibfield  {author} {\bibinfo {author} {\bibfnamefont {F.~G.}\ \bibnamefont {Brandao}}\ and\ \bibinfo {author} {\bibfnamefont {M.~J.}\ \bibnamefont {Kastoryano}},\ }\bibfield  {title} {\bibinfo {title} {Finite correlation length implies efficient preparation of quantum thermal states},\ }\href@noop {} {\bibfield  {journal} {\bibinfo  {journal} {Communications in Mathematical Physics}\ }\textbf {\bibinfo {volume} {365}},\ \bibinfo {pages} {1} (\bibinfo {year} {2019})}\BibitemShut {NoStop}%
\bibitem [{\citenamefont {Gondolf}\ \emph {et~al.}(2024)\citenamefont {Gondolf}, \citenamefont {Scalet}, \citenamefont {Ruiz-de Alarcon}, \citenamefont {Alhambra},\ and\ \citenamefont {Capel}}]{gondolf2024conditional}%
  \BibitemOpen
  \bibfield  {author} {\bibinfo {author} {\bibfnamefont {P.}~\bibnamefont {Gondolf}}, \bibinfo {author} {\bibfnamefont {S.~O.}\ \bibnamefont {Scalet}}, \bibinfo {author} {\bibfnamefont {A.}~\bibnamefont {Ruiz-de Alarcon}}, \bibinfo {author} {\bibfnamefont {A.~M.}\ \bibnamefont {Alhambra}},\ and\ \bibinfo {author} {\bibfnamefont {A.}~\bibnamefont {Capel}},\ }\bibfield  {title} {\bibinfo {title} {Conditional independence of 1d gibbs states with applications to efficient learning},\ }\href@noop {} {\bibfield  {journal} {\bibinfo  {journal} {arXiv preprint arXiv:2402.18500}\ } (\bibinfo {year} {2024})}\BibitemShut {NoStop}%
\bibitem [{\citenamefont {Junge}\ \emph {et~al.}(2018)\citenamefont {Junge}, \citenamefont {Renner}, \citenamefont {Sutter}, \citenamefont {Wilde},\ and\ \citenamefont {Winter}}]{junge2018universal}%
  \BibitemOpen
  \bibfield  {author} {\bibinfo {author} {\bibfnamefont {M.}~\bibnamefont {Junge}}, \bibinfo {author} {\bibfnamefont {R.}~\bibnamefont {Renner}}, \bibinfo {author} {\bibfnamefont {D.}~\bibnamefont {Sutter}}, \bibinfo {author} {\bibfnamefont {M.~M.}\ \bibnamefont {Wilde}},\ and\ \bibinfo {author} {\bibfnamefont {A.}~\bibnamefont {Winter}},\ }\bibfield  {title} {\bibinfo {title} {Universal recovery maps and approximate sufficiency of quantum relative entropy},\ }in\ \href@noop {} {\emph {\bibinfo {booktitle} {Annales Henri Poincar{\'e}}}},\ Vol.~\bibinfo {volume} {19}\ (\bibinfo {organization} {Springer},\ \bibinfo {year} {2018})\ pp.\ \bibinfo {pages} {2955--2978}\BibitemShut {NoStop}%
\bibitem [{\citenamefont {Dennis}\ \emph {et~al.}(2002)\citenamefont {Dennis}, \citenamefont {Kitaev}, \citenamefont {Landahl},\ and\ \citenamefont {Preskill}}]{dennis2002topological}%
  \BibitemOpen
  \bibfield  {author} {\bibinfo {author} {\bibfnamefont {E.}~\bibnamefont {Dennis}}, \bibinfo {author} {\bibfnamefont {A.}~\bibnamefont {Kitaev}}, \bibinfo {author} {\bibfnamefont {A.}~\bibnamefont {Landahl}},\ and\ \bibinfo {author} {\bibfnamefont {J.}~\bibnamefont {Preskill}},\ }\bibfield  {title} {\bibinfo {title} {Topological quantum memory},\ }\href@noop {} {\bibfield  {journal} {\bibinfo  {journal} {Journal of Mathematical Physics}\ }\textbf {\bibinfo {volume} {43}},\ \bibinfo {pages} {4452} (\bibinfo {year} {2002})}\BibitemShut {NoStop}%
\bibitem [{\citenamefont {Lu}\ \emph {et~al.}(2020)\citenamefont {Lu}, \citenamefont {Hsieh},\ and\ \citenamefont {Grover}}]{neg}%
  \BibitemOpen
  \bibfield  {author} {\bibinfo {author} {\bibfnamefont {T.-C.}\ \bibnamefont {Lu}}, \bibinfo {author} {\bibfnamefont {T.~H.}\ \bibnamefont {Hsieh}},\ and\ \bibinfo {author} {\bibfnamefont {T.}~\bibnamefont {Grover}},\ }\bibfield  {title} {\bibinfo {title} {Detecting topological order at finite temperature using entanglement negativity},\ }\href {https://doi.org/10.1103/PhysRevLett.125.116801} {\bibfield  {journal} {\bibinfo  {journal} {Phys. Rev. Lett.}\ }\textbf {\bibinfo {volume} {125}},\ \bibinfo {pages} {116801} (\bibinfo {year} {2020})}\BibitemShut {NoStop}%
\bibitem [{Note1()}]{Note1}%
  \BibitemOpen
  \bibinfo {note} {All the logarithms in this work, including those show up in the definition of entropic quantities, use $2$ as the base.}\BibitemShut {Stop}%
\bibitem [{\citenamefont {Kwon}\ \emph {et~al.}(2022)\citenamefont {Kwon}, \citenamefont {Mukherjee},\ and\ \citenamefont {Kim}}]{kwon2022reversing}%
  \BibitemOpen
  \bibfield  {author} {\bibinfo {author} {\bibfnamefont {H.}~\bibnamefont {Kwon}}, \bibinfo {author} {\bibfnamefont {R.}~\bibnamefont {Mukherjee}},\ and\ \bibinfo {author} {\bibfnamefont {M.}~\bibnamefont {Kim}},\ }\bibfield  {title} {\bibinfo {title} {Reversing lindblad dynamics via continuous petz recovery map},\ }\href@noop {} {\bibfield  {journal} {\bibinfo  {journal} {Physical Review Letters}\ }\textbf {\bibinfo {volume} {128}},\ \bibinfo {pages} {020403} (\bibinfo {year} {2022})}\BibitemShut {NoStop}%
\bibitem [{Note2()}]{Note2}%
  \BibitemOpen
  \bibinfo {note} {Naively the evolution time of the reversed dynamics {$\setbox \z@ \hbox {\mathsurround \z@ $\protect \textstyle \protect \mathcal {G}$}\mathaccent "0365{\protect \mathcal {G}}$} is {$O(r^{D})$} because the circuit depth is multiplied by the same factor due to the reorganization. However it can be turned into a time $1$ evolution by absorbing the factor into {$\protect \mathcal {L}(\tau )$}.}\BibitemShut {Stop}%
\bibitem [{\citenamefont {Murg}\ \emph {et~al.}(2007)\citenamefont {Murg}, \citenamefont {Verstraete},\ and\ \citenamefont {Cirac}}]{Murg2007peps}%
  \BibitemOpen
  \bibfield  {author} {\bibinfo {author} {\bibfnamefont {V.}~\bibnamefont {Murg}}, \bibinfo {author} {\bibfnamefont {F.}~\bibnamefont {Verstraete}},\ and\ \bibinfo {author} {\bibfnamefont {J.~I.}\ \bibnamefont {Cirac}},\ }\bibfield  {title} {\bibinfo {title} {Variational study of hard-core bosons in a two-dimensional optical lattice using projected entangled pair states},\ }\href {https://doi.org/10.1103/PhysRevA.75.033605} {\bibfield  {journal} {\bibinfo  {journal} {Phys. Rev. A}\ }\textbf {\bibinfo {volume} {75}},\ \bibinfo {pages} {033605} (\bibinfo {year} {2007})}\BibitemShut {NoStop}%
\bibitem [{\citenamefont {Bravyi}\ \emph {et~al.}(2014)\citenamefont {Bravyi}, \citenamefont {Suchara},\ and\ \citenamefont {Vargo}}]{PhysRevA.90.032326}%
  \BibitemOpen
  \bibfield  {author} {\bibinfo {author} {\bibfnamefont {S.}~\bibnamefont {Bravyi}}, \bibinfo {author} {\bibfnamefont {M.}~\bibnamefont {Suchara}},\ and\ \bibinfo {author} {\bibfnamefont {A.}~\bibnamefont {Vargo}},\ }\bibfield  {title} {\bibinfo {title} {Efficient algorithms for maximum likelihood decoding in the surface code},\ }\href {https://doi.org/10.1103/PhysRevA.90.032326} {\bibfield  {journal} {\bibinfo  {journal} {Phys. Rev. A}\ }\textbf {\bibinfo {volume} {90}},\ \bibinfo {pages} {032326} (\bibinfo {year} {2014})}\BibitemShut {NoStop}%
\bibitem [{Note3()}]{Note3}%
  \BibitemOpen
  \bibinfo {note} {We remark that the {$p=0.5$} state requires infinite time dephasing Lindbladian evolution. But as we show in SM, a $O(\log L)$ time evolution is enough to obtain a sufficiently close-by state}\BibitemShut {NoStop}%
\bibitem [{\citenamefont {Ozeki}\ and\ \citenamefont {Nishimori}(1993)}]{ozeki1993phase}%
  \BibitemOpen
  \bibfield  {author} {\bibinfo {author} {\bibfnamefont {Y.}~\bibnamefont {Ozeki}}\ and\ \bibinfo {author} {\bibfnamefont {H.}~\bibnamefont {Nishimori}},\ }\bibfield  {title} {\bibinfo {title} {Phase diagram of gauge glasses},\ }\href@noop {} {\bibfield  {journal} {\bibinfo  {journal} {Journal of Physics A: Mathematical and General}\ }\textbf {\bibinfo {volume} {26}},\ \bibinfo {pages} {3399} (\bibinfo {year} {1993})}\BibitemShut {NoStop}%
\bibitem [{\citenamefont {Merz}\ and\ \citenamefont {Chalker}(2002)}]{merz2002rbim}%
  \BibitemOpen
  \bibfield  {author} {\bibinfo {author} {\bibfnamefont {F.}~\bibnamefont {Merz}}\ and\ \bibinfo {author} {\bibfnamefont {J.}~\bibnamefont {Chalker}},\ }\bibfield  {title} {\bibinfo {title} {Two-dimensional random-bond ising model, free fermions, and the network model},\ }\href@noop {} {\bibfield  {journal} {\bibinfo  {journal} {Physical Review B}\ }\textbf {\bibinfo {volume} {65}},\ \bibinfo {pages} {054425} (\bibinfo {year} {2002})}\BibitemShut {NoStop}%
\bibitem [{\citenamefont {Leifer}\ and\ \citenamefont {Poulin}(2008)}]{leifer2008quantum}%
  \BibitemOpen
  \bibfield  {author} {\bibinfo {author} {\bibfnamefont {M.~S.}\ \bibnamefont {Leifer}}\ and\ \bibinfo {author} {\bibfnamefont {D.}~\bibnamefont {Poulin}},\ }\bibfield  {title} {\bibinfo {title} {Quantum graphical models and belief propagation},\ }\href@noop {} {\bibfield  {journal} {\bibinfo  {journal} {Annals of Physics}\ }\textbf {\bibinfo {volume} {323}},\ \bibinfo {pages} {1899} (\bibinfo {year} {2008})}\BibitemShut {NoStop}%
\bibitem [{\citenamefont {Kuwahara}\ \emph {et~al.}(2020)\citenamefont {Kuwahara}, \citenamefont {Kato},\ and\ \citenamefont {Brand\~ao}}]{tomotaka2020clustering}%
  \BibitemOpen
  \bibfield  {author} {\bibinfo {author} {\bibfnamefont {T.}~\bibnamefont {Kuwahara}}, \bibinfo {author} {\bibfnamefont {K.}~\bibnamefont {Kato}},\ and\ \bibinfo {author} {\bibfnamefont {F.~G. S.~L.}\ \bibnamefont {Brand\~ao}},\ }\bibfield  {title} {\bibinfo {title} {Clustering of conditional mutual information for quantum gibbs states above a threshold temperature},\ }\href {https://doi.org/10.1103/PhysRevLett.124.220601} {\bibfield  {journal} {\bibinfo  {journal} {Phys. Rev. Lett.}\ }\textbf {\bibinfo {volume} {124}},\ \bibinfo {pages} {220601} (\bibinfo {year} {2020})}\BibitemShut {NoStop}%
\bibitem [{\citenamefont {Kato}\ and\ \citenamefont {Brandao}(2019)}]{kato2019quantum}%
  \BibitemOpen
  \bibfield  {author} {\bibinfo {author} {\bibfnamefont {K.}~\bibnamefont {Kato}}\ and\ \bibinfo {author} {\bibfnamefont {F.~G.}\ \bibnamefont {Brandao}},\ }\bibfield  {title} {\bibinfo {title} {Quantum approximate markov chains are thermal},\ }\href@noop {} {\bibfield  {journal} {\bibinfo  {journal} {Communications in Mathematical Physics}\ }\textbf {\bibinfo {volume} {370}},\ \bibinfo {pages} {117} (\bibinfo {year} {2019})}\BibitemShut {NoStop}%
\end{thebibliography}%
 
\appendix
\onecolumngrid
\section{Petz recovery map}
Let $\calH$, $\calH'$ be two Hilbert spaces. For a density operator $\rho$ defined on $\calH$ and a completely positive trace preserving (CPTP) map $\calE:\calB(\calH)\rightarrow\calB(\calH')$, the CPTP map called twirled Petz map $\calP_{\calE,\rho}: \calB(\calH')\rightarrow\calB(\calH)$ is defined as:
\begin{equation}
    \calP_{\calE,\rho}[\cdot] = 
    \int_{-\infty}^{\infty}
    f(\tau)
    \rho^{\frac{1 - i\tau}{2}} 
    \calE^\dagger\left[ 
    \calE[\rho]^{\frac{-1 + i\tau}{2}} 
    (\cdot) 
    \calE[\rho]^{\frac{-1 - i\tau}{2}} 
    \right] 
    \rho^{\frac{1 + i\tau}{2}} 
    \d\tau 
\end{equation}
where $f(t)\equiv\frac{\pi}{2(\cosh(\pi t) + 1)}$, and the map $\calE^\dagger$ is defined through the relation $\tr(\calE^\dagger[X] Y)=\tr(X \calE[Y])$

The importance of the twirled Petz map comes from the following inequality concerning approximate quantum sufficiency~\cite{junge2018universal}:
\begin{equation}
    D(\rho \|\sigma) - D(\calE[\rho] \| \calE[\sigma]) \geq -2\log F(\rho, \calP_{\calE,\sigma}\circ\calE[\rho])
\end{equation}
Where $\rho$ and $\sigma$ are both states defined on $\calH$. $D(\rho\|\sigma)\equiv\tr(\rho(\log\rho-\log\sigma))$ and $F(\cdot,\cdot)$ is the fidelity between states. In plain language, the theorem states that the twirled Petz map $\calP_{\calE,\sigma}$ approximately reverses the action of $\calE$ on any state $\rho$ whose relative entropy with respect to $\sigma$ does not change much under $\calE$.

In order to derive Eq.\eqref{eq: single_err_bound}, we take $\rho=\rho_{ABC}$ and $\sigma = \rho_{AB}\otimes\rho_C$, and let $\mathcal{E}$ be a channel acting on $A$ only. In this case the \textit{l.h.s.} becomes:
\begin{equation}
    I_\rho(AB:C) - I_{\calE[\rho]}(AB:C)=I_\rho(A:C|B) - I_{\calE[\rho]}(A:C|B)
\end{equation}
which comes from $I(A:C|B)\equiv I(AB:C)-I(B:C)$. We notice that $I_{\rho}(B:C)=I_{\calE[\rho]}(B:C)$ because $\calE$ acts on $A$ only. To bound the \textit{r.h.s.} in terms of trace norm, we use the relation:
\begin{equation}
    1-F(\rho,\sigma)\geq \frac{1}{4}|\rho-\sigma|^2_1
    ~~~\Rightarrow~~~
    -2\log F \geq \frac{1}{2\ln2}|\rho-\sigma|^2_1
\end{equation}
Combining two expressions together, we obtain:
\begin{equation}
    I_\rho(A:C|B) - I_{\calE[\rho]}(A:C|B) \geq (2\ln2)^{-1} \cdot| \calP_{\calE,\rho_{AB}}\circ\calE[\rho] - \rho|_1^2
\end{equation}
which is also Eq.\eqref{eq: single_err_bound}.

\section{Derivation of Eq.\eqref{eq: total_err_bound}}
The recovery error of a single layer is bounded as:
\begin{equation}
\begin{aligned}
    \left|\tilde{\calC}_\ell\circ{\calC}_\ell[\rho_{\ell-1}] - \rho_{\ell-1}\right|_1
    &=
    \left|
    \sum_{x=1}^{x_{\rm max}-1}
    \left(
    \tilde{\calE}_{x, \ell}\circ\calE_{x, \ell}\left[\calC_{<x, \ell}[\rho_{\ell-1}]\right] - \calC_{<x, \ell}[\rho_{\ell-1}]
    \right)
    \right|_1\\
    &\leq
    \sum_{x=1}^{x_{\rm max}-1}
    \left|
    \tilde{\calE}_{x, \ell}\circ\calE_{x, \ell}\left[\calC_{<x, \ell}[\rho_{\ell-1}]\right] - \calC_{<x, \ell}[\rho_{\ell-1}]
    \right|_1\\
    &\leq
    \sum_{x=1}^{x_{\rm max}-1}
    \left|
    \tilde{\calE}_{x, \ell}\circ\calE_{x, \ell}\left[\rho_{\ell-1}\right] - \rho_{\ell-1}
    \right|_1\\
\end{aligned}
\end{equation}
where $\calC_{<x, \ell}\equiv\prod_{y<x}\tilde{\calE}_{y,\ell}\circ\calE_{y,\ell}$. The first inequality is from the triangle inequality of trace norm, while the second one is from the contractivity of CPTP maps. 

The cumulative error of the total forward-backward evolution is:
\begin{equation}
\left|\tilde{\calG}(t)\circ{\calG}(t)[\rho] - \rho\right|_1
     \equiv \left|\tilde{\calC}_{1}\circ...\circ\tilde{\calC}_{\ell_{\rm max}}\circ\calC_{\ell_{\rm max}}\circ...\circ\calC_{1}[\rho] - \rho\right|_1
     \equiv \left|\tilde{\calC}_{\{1,...,\ell_{\rm max}\}}[\rho_{\ell_{\rm max}}] - \rho_0\right|_1
\end{equation}
where $\ell_{\rm max}=t/\delta t$, and $\tilde{\calC}_{\{1,...,\ell\}}\equiv\tilde{\calC}_{1}\circ...\circ\tilde{\calC}_{\ell}$. We make use of the following iteration relation which holds for $\ell=1,...\ell_{\rm max}$:
\begin{equation}
    \tilde{\calC}_{\{1,...,\ell\}}[\rho_{\ell}] = \tilde{\calC}_{\{1,...,\ell-1\}}[\rho_{\ell-1}] + 
    \tilde{\calC}_{\{1,...,\ell-1\}}\left[\tilde{\calC}_\ell\circ\calC_\ell[\rho_{\ell-1}]-\rho_{\ell-1}\right]
\end{equation}
so that the cumulative error can be expanded and bounded as follows:
\begin{equation}
    \begin{aligned}
        \left|\tilde{\calC}_{\{1,...,\ell_{\rm max}\}}[\rho_{\ell_{\rm max}}] - \rho_0\right|_1 
        &=
        \left|
        \sum_{\ell=1}^{\ell_{\rm max}-1}
    \tilde{\calC}_{\{1,...,\ell-1\}}\left[\tilde{\calC}_\ell\circ\calC_\ell[\rho_{\ell-1}]-\rho_{\ell-1}\right]
        \right|_1\\
        &\leq
        \sum_{\ell=1}^{\ell_{\rm max}-1}
        \left|
        \tilde{\calC}_{\{1,...,\ell-1\}}\left[\tilde{\calC}_\ell\circ\calC_\ell[\rho_{\ell-1}]-\rho_{\ell-1}\right]
        \right|_1\\
        &\leq
        \sum_{\ell=1}^{\ell_{\rm max}-1}
        \left|
        \tilde{\calC}_\ell\circ\calC_\ell[\rho_{\ell-1}]-\rho_{\ell-1}
        \right|_1\\
        &\leq \sum_{x, \ell}
    \left|
    \tilde{\calE}_{x, \ell}\circ\calE_{x, \ell}\left[\rho_{\ell-1}\right] - \rho_{\ell-1}
    \right|_1
    \end{aligned}
\end{equation}

\section{Derivation of Eq.\eqref{eq: vn_to_shannon}}
Suppose $Q$ is a non-simply-connected region of the toric code ground states surrounding a hole.
\begin{figure}[h!]
    \centering
    \includegraphics[width=0.20\linewidth]{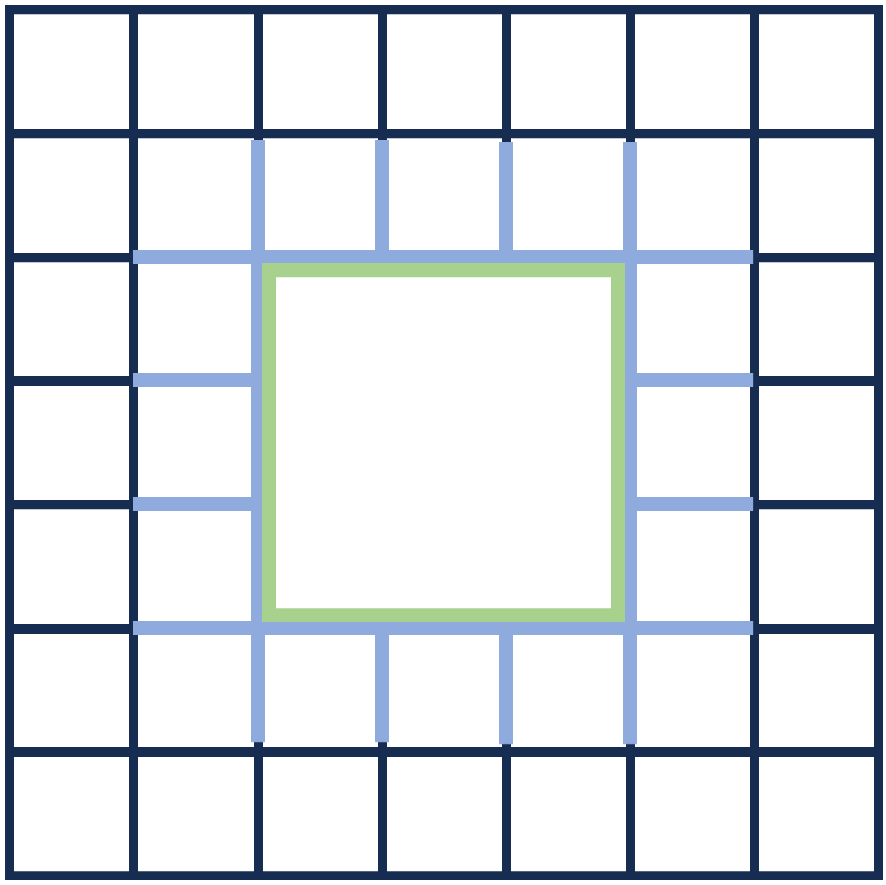}
    \caption{Illustration of a non-simply connected region Q. Only qubits (edges) that belong to $Q$ is drawn. Supports of the two non-local operators $A_{\tilde\square}$ and $B_{\tilde +}$ surrounding the hole are denoted with green and blue edges, respectively.}
    \label{apfig: lattice}
\end{figure}

Before the dephasing, the reduced density operator on $Q$ is:
\begin{equation}
\begin{aligned}
    \rho_{0, Q} &=\tr\left(\ket{\rm t.c.}\bra{\rm t.c.}\right) \\
    &= 
    \left(\frac{1+A_{\tilde\square}}{2}\right)
    \left(\frac{1+B_{\tilde+}}{2}\right)
    \prod_{\square\ {\rm within}\ Q}\left(\frac{1+A_\square}{2}\right) \prod_{+\ {\rm within}\ Q} \left(\frac{1+B_+}{2}\right)
\end{aligned}
\end{equation}
where $A_{\tilde\square}$ is a $X$-loop operator acting on green edges (see Fig.\ref{apfig: lattice}), and $B_{\tilde +}$ is a $Z$-loop operator acting on blue edges. The two terms show up because $Q$ is not simply-connected. We notice that each factor in the expression is a projector operator and all factors commute with each other.

Once a $Z$ operator acts on an edge, anyon occupancies of the two adjacent plaquettes will be flipped. Thus if we use the binary vector $\be$ to indicate the set of edges that are acted by $Z$,  plaquettes with a net anyon (indicated with a binary vector $\bm$) are those intersects odd number of times with $\be$. We denote this relation as $\bm=\partial\be$. The dephased state is the weighted mixture of states result from all the possible $\be$s:
\begin{equation}
\begin{aligned}
    \rho_{p, Q} 
    &= \frac{1}{2^{z_Q}} \sum_{\bm}\Pr(\bm)
    \left(\frac{1+B_{\tilde+}}{2}\right)
    \left(\frac{1+(-1)^{m_{\tilde\square}}A_{\tilde\square}}{2}\right)
    \prod_{\square\ {\rm within}\ Q}
    \left(\frac{1+(-1)^{m_{\square}} A_\square}{2}\right)
    \prod_{+\ {\rm within}\ Q}
    \left(\frac{1+ B_+}{2}\right)\\
    &\equiv \frac{1}{2^{z_Q}} \sum_{\bm}\Pr(\bm)\Pi_{\bm}
\end{aligned}
\end{equation}
where $\Pr(\bm) = \sum_{\be}p^{|\be|}(1-p)^{|Q|-|\be|}\delta(\partial\be=\bm)$. Noticing that each $\Pi_{\bm}$ is a projector and $\Pi_{\bm}\Pi_{\bm'}=0$ when $\bm\neq\bm'$, we obtain the expression for $\rho_{p,Q}$' s  von Neumann entropy:
\begin{equation}
    S(\rho_{p, Q}) = -\tr(\rho_{p, Q} \log\rho_{p, Q}) = S(\rho_{0, Q}) + H(\bm)
\end{equation}
Which is the Eq.\eqref{eq: vn_to_shannon} in the maintext. We remark that there is a small difference between the notation here and that adapted in the maintext: In the maintext $\bm$ represents only anyon configuration of unit plaquettes within $Q$, and the net anyon number in the big plaquette (\textit{i.e.} the green plaquette above) is denoted with $\pi(\bm_\Gamma)$. But here we use $\bm$ to denote both.

\section{Tensor network technique for simulating $H(\bm)$}
In order to simulate $H(\bm)$, we first rewrite it as a sample averaged quantity:
\begin{equation}
    H(\bm) = -\mathbb{E}_{\bm\sim\Pr(\bm)}\left[\log\Pr(\bm)\right].
\end{equation}
Anyon configuration $\bm$ can be efficiently sampled by first sampling $\be$, which follows a product distribution, then calculating $\bm$ as $\bm=\partial\be$. However a bruteforce evaluation of $\Pr(\bm)$ is not easy because it requires enumerating all the $\be$s that can produce $\bm$, which leads to exponentially many terms.

To circumvant this difficulty, we represent $\Pr(\bm)$ as a two dimensional tensor network. For instance, for the following anyon configuration $\bm$ (plaquattes holding anyons are shaded):
\begin{equation}
    \eqfig{3.5cm}{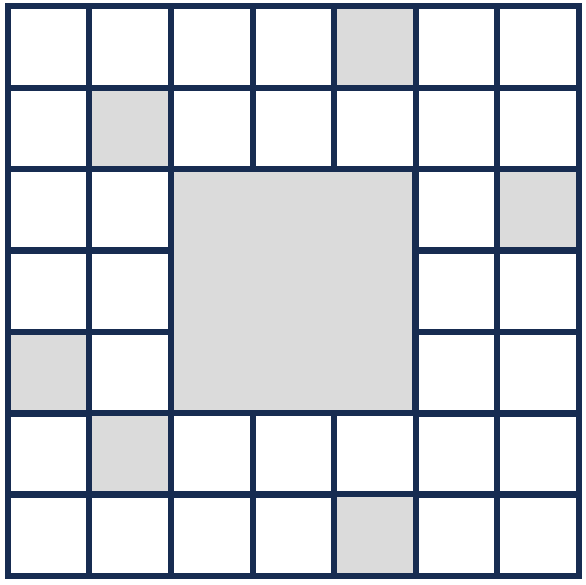},
\end{equation}
its probability can be expressed as:
\begin{equation}
    \Pr(\bm)=\sum_{\be}p^{|\be|}(1-p)^{|Q|-|\be|}\delta(\partial\be=\bm)
    =\eqfig{4cm}{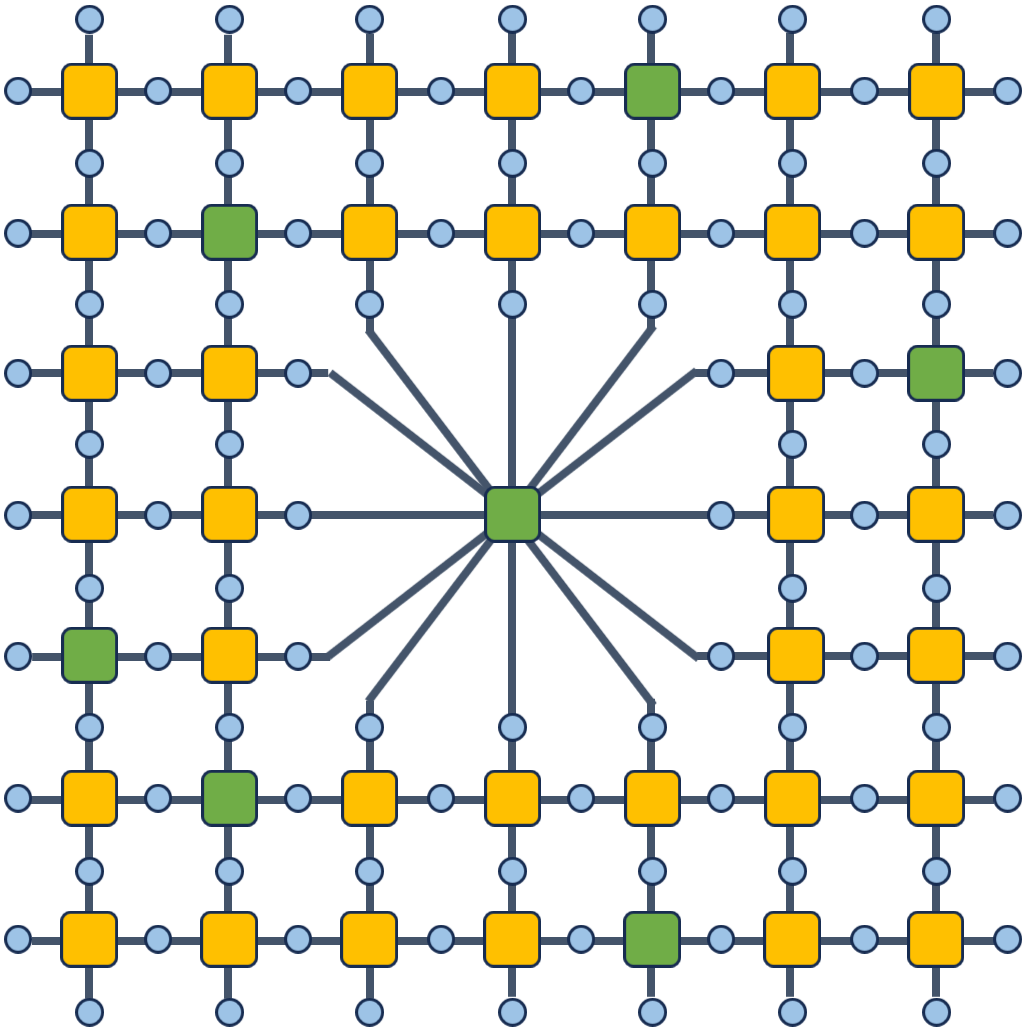}
    =\eqfig{4cm}{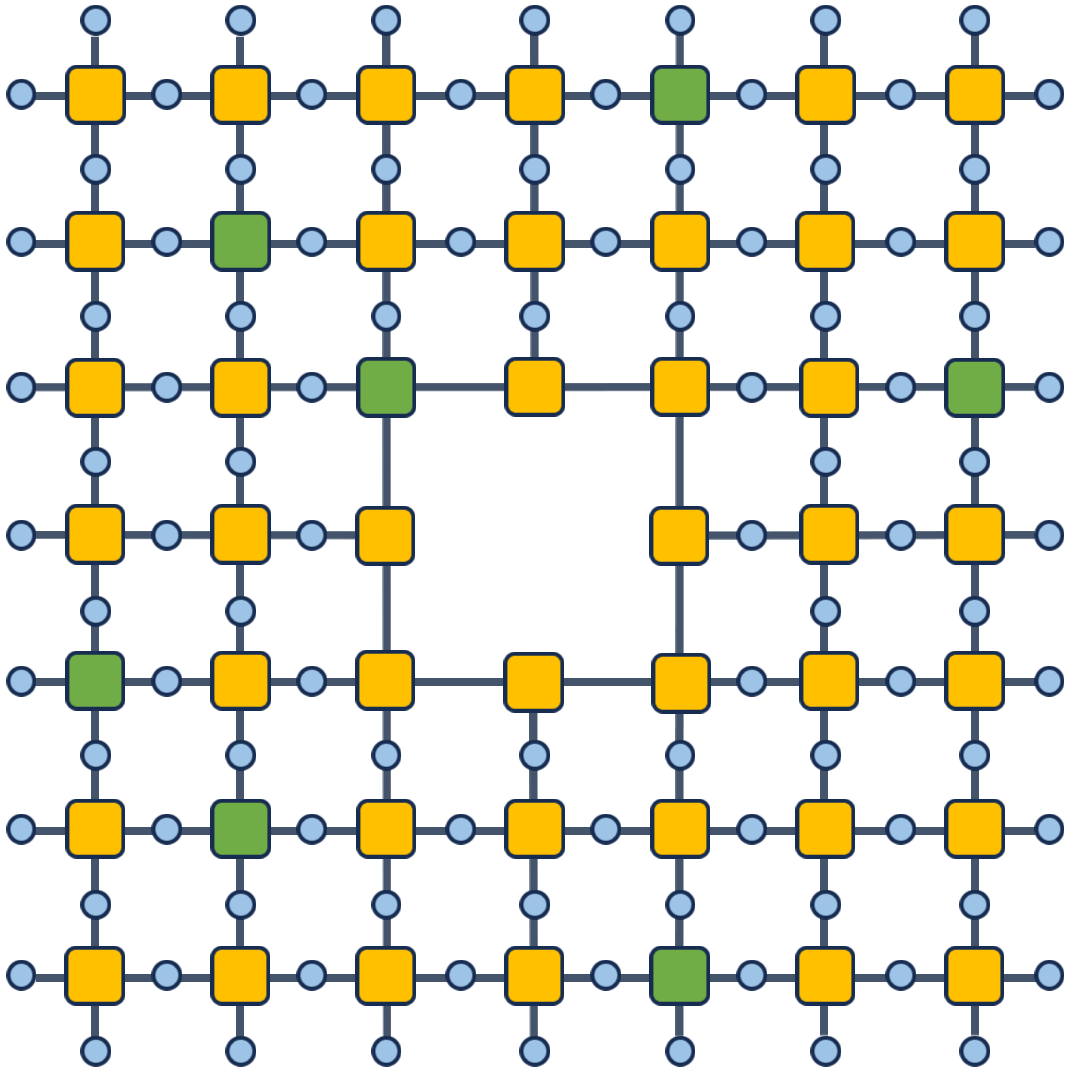}
    ,
\end{equation}
where each two-leg circle tensor $T_{s_1 s_2},\ s_{1,2}\in\{0,1\}$ is used for assigning weights:
\begin{equation}
    T_{s_1 s_2} = \delta(s1=s2)p^{s_1}(1-p)^{s_1}
\end{equation}
and each $q$-leg square tensor $Q^s_{s_1,...,s_q},\ s_{i}\in\{0,1\}$ is used for imposing parity constraint on each plaquatte:
\begin{equation}
    Q^s_{s_1,...,s_q} = \delta\left(\sum_{i=1}^q s_i=s\mod 2\right).
\end{equation}

In the tensor network image above, $Q^0$s and $Q^1$s are drawn with yellow and green squares, respectively. Correctness of the tensor network representation can be explicitly checked by expanding all the tensors. The 2D tensor network can be evaluated approximately and efficiently using the boundary matrix product state (bMPS) method~\cite{Murg2007peps, PhysRevA.90.032326}. 

\section{Details on numerical simulation in Fig.\ref{fig: numerics}}
For numerical results presented in Fig.\ref{fig: numerics}, regions $A$, $B$ and $C$ are taken as follows (the plotted figure correponds to $r=2$):
\begin{equation*}
    \eqfig{5cm}{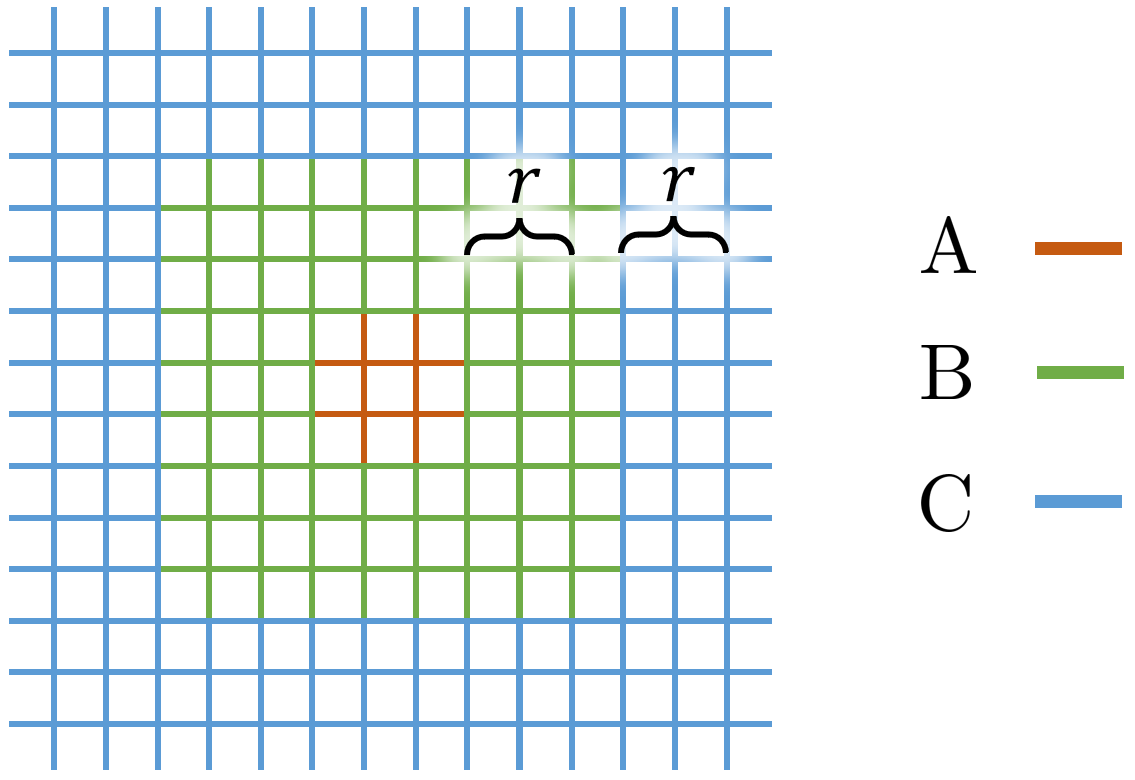}.
\end{equation*}
Edges belong to different regions are indicated with different colors. When varrying $r$, the region $A$ remains unchanged. 

In Figs.\ref{fig: numerics}(b, c), each data point is averaged over at least $3.5\times10^4$ samples. In Fig.\ref{fig: numerics}(d), each data point is averaged over at least $6\times10^6$ samples. 

\section{Mapping to RBIM's free energy}
In this appendix we derive the mapping from the anyon distribution's Shannon entropy $H(\bm_Q)$ (\textit{r.h.s.} of Eq.\eqref{eq: vn_to_shannon}) to the RBIM's free energy, and further relate CMI to RBIM's free energy cost due to a point defect.

We focus on an annulus shaped subregion $Q$ which contains a hole. $\bm_Q$ is used for denoting both the anyon configuration within $Q$ and the net anyon occupancy in the hole, \textit{i.e.} the hole is treated as a big plaqquette. A simply connected region $Q$ can be considered as a special case where the hole contains only one plaquette.

We use binary vector $\be$ to indicate edges that are acted by $Z$ gates, and $\bm=\partial\be$ is its corresponding anyon configuration. The probability of observing the configuration $\bm$ is:
\begin{equation}
        \Pr(\bm)
        =
        \sum_{\be'}\delta(\partial\be=\partial\be') p^{|\be'|}(1-p)^{|Q|-|\be'|}.
\end{equation}
After a change of variable: $\bc \equiv\be+\be'$ (the summation is modular 2), the delta function constraint becomes into $\partial\bc=0$, \textit{i.e.} $\bc$ must form a loop. We obtain:
\begin{equation}
        \Pr(\bm) = \left[p(1-p)\right]^{|Q|/2} \sum_{\bc}\delta(\partial\bc=0) e^{J\sum_x\eta_x \tilde{c}_x}
\end{equation}
where $\tilde{c}_x \equiv 2c_x-1$, $\tilde{e}_x \equiv 2e_x-1$, $J=\frac{1}{2}\log(p/(1-p))$, and the index $x$ runs over edges. 

In order to solve the loop constraint, we perform another change of variable: $c_{x=ij}=\sigma_i\sigma_j$, where $\sigma$s are spin variables defined on plaquettes of the orignal lattice, \textit{i.e.} the dual lattice:
\begin{equation}
    \Pr(\bm) = \frac{1}{2}\left[p(1-p)\right]^{|Q|/2} \sum_{\{\sigma\}} e^{J\sum_{ij} \eta_{ij}\sigma_i\sigma_j} = e^{-F_{{\rm RBIM},p}(Q, \tilde{\be})-c_1|Q|-c_2}
\end{equation}
where $c_1 = -\frac{1}{2}\ln p(1-p)$ and $c_2=\ln 2$. 

The Shannon entropy of $\bm$ can now be written as:
\begin{align}
    H(\bm_Q) 
    &= -\sum_{\bm_Q} \Pr(\bm_Q)\log\Pr(\bm)\\
    &= \sum_{\bm_Q} \Pr(\bm_Q)(F_{{\rm RBIM},p}(Q, \tilde{\be})+c_1|Q|+c_2)\\
    &= \sum_{\be} \Pr(\be)(F_{{\rm RBIM},p}(Q, \tilde{\be})+c_1|Q|+c_2)\\
    &= \overline{F_{{\rm RBIM}, p}(Q)}+c_1|Q|+c_2
\end{align}

Finally we look into the meaning of CMI in terms of the RBIM. Letting $Q=B$  and $Q=AB$ following the geometry below:
\begin{equation*}
    \eqfig{2.8cm}{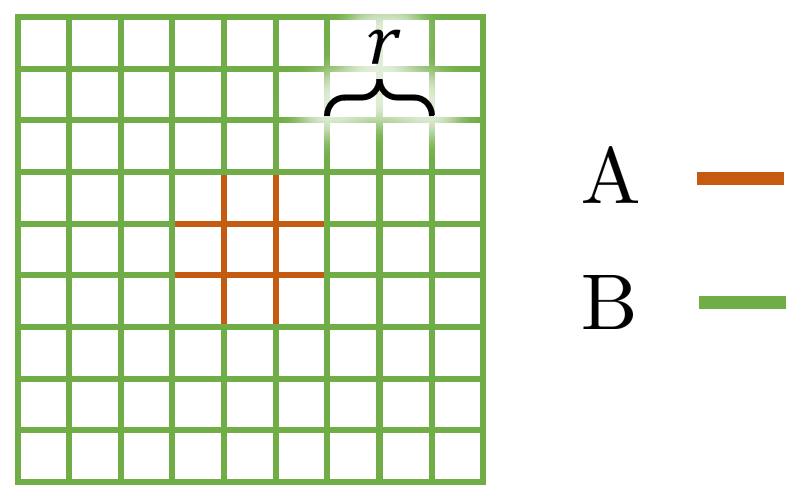}
\end{equation*}
we obtain that:
\begin{align}
    &H(\pi(\bm_A), \bm_B) - H(\bm_{AB})\\
    =& 
    \overline{F_{{\rm RBIM}, p}}\left(\eqfig{2.8cm}{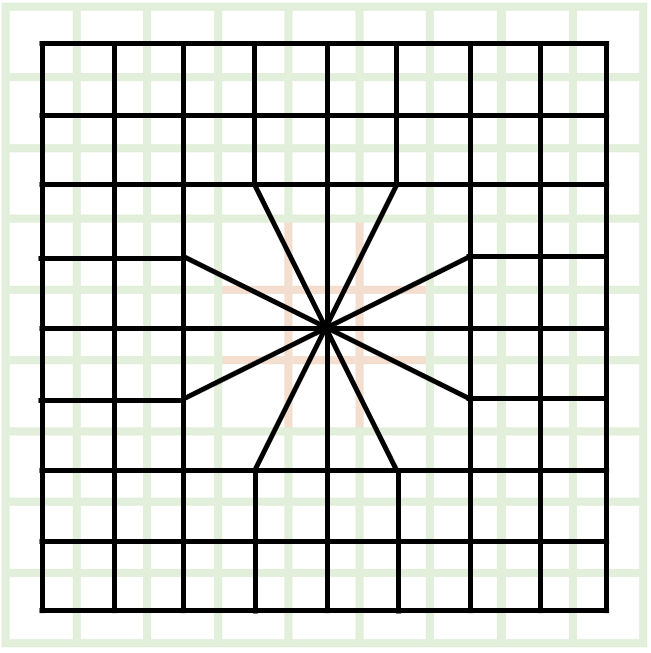}\right)
    -
    \overline{F_{{\rm RBIM}, p}}\left(\eqfig{2.8cm}{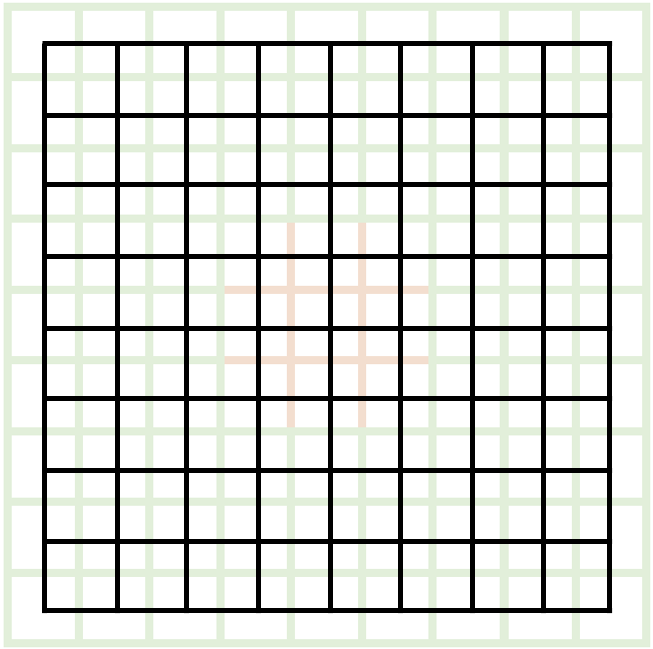}\right) + c.c.
\end{align}
where lattices upon which the correponding RBIMs are defined were drawn. Since $A$ is fixed, in the large $r$ limit the expression above can be viewed as the free energy cost of introducing a point-like defect in the center of the lattice, which we denote with $F_{\rm def}(2r)$. Following exactly the same argument, one can derive that $H(\pi(\bm_A), \bm_{BC}) - H(\bm_{ABC})=F_{\rm def}(4r)$. We thus obtain the Eq.\eqref{eq: F_defect}:
\begin{align}
    I(A:C|B) 
    &= H(\pi(\bm_A), \bm_{BC}) - H(\bm_{ABC}) - (H(\pi(\bm_A), \bm_{B}) - H(\bm_{AB}))\\
    &= \overline{F_{\rm def}(4r)} - \overline{F_{\rm def}(2r)},
\end{align}

\section{Convergence of dephasing Lindbladian}
In this appendix we examine how fast does $\exp(t\calL)$, where $\calL[\cdot]\equiv\sum_i\calL_i=\sum_{i}\frac{1}{2}(\cdot)-\frac{1}{2}Z_i(\cdot)Z_i$, converges to the complete dephasing channel $\prod_i\calE_i[\cdot]\equiv\prod_i\left(\frac{1}{2}(\cdot)+\frac{1}{2}Z_i(\cdot)Z_i\right)$. 

We consider the diamond distance: for two quantum channels $\calC_1$, $\calC_2$ acting on a $n$-dimensional Hilbert space, their diamond distance is defined as
\begin{equation}
    \left|\calC_1-\calC_2\right|_\diamond \equiv 
    \max_{\rho} \left|\calC_1\otimes \mathcal{I}_n[\rho]-\calC_2\otimes \mathcal{I}_n[\rho]\right|_1
\end{equation}
where $\mathcal{I}_n$ is the identity channel. 

We first check the distance between $\exp(\calL_i)$ and $\calE_i$. Bringing in $\calC_1 =\exp(t\calL_i)$ and $\calC_2=\calE_i$ , we get:
\begin{equation}
    \left|\exp(t\calL_i)-\calE_i\right|_\diamond
    = \left|p_t-\frac{1}{2}\right| \cdot \max_{\rho} \left|Z\rho Z - \rho\right|_1
    = \frac{1}{2}e^{-t}\lambda
\end{equation}
where $p_t = (1-e^{-t})/2$. We let $\lambda \equiv \max_{\rho} \left|Z\rho Z - \rho\right|_1$, which is an $O(1)$ constant.

Then the diamond distance between $\calE$ and $\exp(t\calL)$ is bounded as:
\begin{equation}
    \left|\exp(t\calL)-\calE\right|_\diamond \leq \sum_i\left|\exp(t\calL_i)-\calE_i\right|_\diamond = \frac{L}{2}e^{-t}\lambda
\end{equation}
Thus in order to acheive $\epsilon$ proximity, it suffices to pick $t=O(\log(L/\epsilon))$. 

In conclusion, $\calE$ can be well-approximated by a quasi-local Lindbladian evolution. 

\end{document}